\documentclass{aa}  

\usepackage{graphicx}
\usepackage{txfonts}
\usepackage{color}
\usepackage{natbib}
\usepackage{colordvi}
\bibpunct{(}{)}{;}{a}{}{,}
\usepackage{soul}

\begin{document}

   \title{
   The limb-brightened jet of M\,87 
   down to 7~Schwarzschild radii scale
   }

   \subtitle{}

   \author{
   J. -Y. Kim \inst{1} 
   \and T. P. Krichbaum \inst{1} 
   \and R. -S. Lu \inst{1}
   \and E. Ros \inst{1,2,3}
   \and U. Bach\inst{1}
   \and M. Bremer \inst{4}  
   \and P. de Vicente \inst{5}
   \and M. Lindqvist \inst{6}
   \and J. A. Zensus \inst{1}
          }

   \institute{
   Max-Planck-Institut f\"ur Radioastronomie, Auf dem H\"ugel 69, 53121 Bonn, Germany
   \\ \email{jykim@mpifr-bonn.mpg.de}
   \and
   Observatori Astron\`omic, Universitat de Val\`encia, Parc Cient\'{\i}fic, C. Catedr\'atico Jos\'e Beltr\'an 2, E-46980 Paterna, Val\`encia, Spain
   \and
   Departament d'Astronomia i Astrof\'{\i}sica, Universitat de Val\`encia, C. Dr. Moliner 50, E-46100 Burjassot, Val\`encia, Spain
   \and       
   Institut de Radio Astronomie Millim\'{e}trique, 300 rue de la Piscine, Domaine Universitaire, Saint Martin d'H\`eres 38406, France
   \and
   Observatorio de Yebes (IGN), Apartado 148, E-19180 Yebes, Spain
   \and
   Department of Space, Earth and Environment, Chalmers University of Technology, Onsala Space Observatory, 439 92 Onsala, Sweden
             }

   \date{Received --; accepted --}

% \abstract{}{}{}{}{} 
% 5 {} token are mandatory
 
  \abstract
   {
   M\,87 is one of the nearest radio galaxies with a prominent jet extending from sub-pc to kpc-scales.
   Because of its proximity and large mass of the central black hole,
   it is one of the best radio sources to study jet formation.
   We aim at studying the physical conditions near the jet base
   at projected separations from the BH of $\sim7-100$ Schwarzschild radii ($R_{\rm sch}$).
   Global mm-VLBI Array (GMVA) observations at 86 GHz ($\lambda=3.5$\,mm) provide an angular resolution of $\sim50\mu$as, 
   which corresponds to a spatial resolution of only $7~R_{\rm sch}$
   and reach the small spatial scale.
   We use five GMVA data sets of M\,87 obtained during 2004--2015 and present new high angular resolution VLBI maps at 86\,GHz.
   In particular, we focus on the analysis of 
   the brightness temperature, the jet ridge lines, and the jet to counter-jet ratio.
   The imaging reveals a parabolically expanding limb-brightened jet which emanates from a resolved VLBI core of $\sim(8-13) R_{\rm sch}$ size.
   The observed brightness temperature of the core at any epoch is $\sim(1-3)\times10^{10}$\,K,
   which is below the equipartition brightness temperature and suggests magnetic energy dominance at the jet base.
   We estimate the diameter of the jet at its base to be 
   $\sim5 R_{\rm sch}$ assuming a self-similar jet structure.
   This suggests that the sheath of the jet may be anchored in the very inner portion of the accretion disk.
   The image stacking reveals faint emission at the center of the edge-brightened jet on sub-pc scales.
   We discuss its physical implication within the context of the spine-sheath structure of the jet.
   }


   \titlerunning{GMVA observations of M\,87}
   \authorrunning{J. -Y. Kim et al.}
               
   \maketitle
%
%-------------------------------------------------------------------

\section{Introduction}\label{sec:intro}

The formation and initial acceleration of relativistic jets in Active Galactic Nuclei (AGN) is still
one of the open fundamental questions in modern astrophysics.
From a theoretical point of view, there is a general agreement that
hot accretion flows falling toward central supermassive black holes (SMBHs) can produce 
collimated streams of highly magnetized plasma, which propagate outwards and are accelerated through 
general magnetohydrodynamic (GRMHD) processes (see \citealt{meier12} and \citealt{yuan14} for a review).
The two most promising mechanisms for jet launching are  
the extraction of rotational energy at the spinning BH ergosphere (e.g., BZ mechanism; \citealt{bz77}) 
and jet launching from a magnetically collimated accretion disk wind (e.g. BP mechanism; \citealt{bp82}), 
which are not mutually exclusive \citep{hardee07}.
Further out, the magnetic energy stored in the jet is gradually converted into kinetic power, 
as the jet expands and  interacts with the jet-ambient medium,
which in turn further collimates and accelerates the jet over long distances
\citep{mckinney06,komissarov07,lyubarsky09,moscibrodzka16}.

Observational constraints on theoretical models are best obtained from imaging of inner jet regions 
through high resolution VLBI imaging.
In this regard and owing to its proximity, the nearby giant elliptical galaxy M\,87 (1228+126, 3C\,274, Virgo\,A) 
is an ideal laboratory for the study of jet launching and the coupling of the jet to the accretion flow and the central BH.
M\,87 is at a distance of $16.7$ Mpc \citep{bird10}, which yields an angular to linear conversion 
scale of 1\,mas $\sim 0.08$~pc. 
Adopting the mass of the SMBH $M_{\rm BH}=6.1\times10^{9}M_{\odot}$ 
(using the mass from \citealt{gebhardt11} adjusted to the distance above)
an angular scale of 1\,mas corresponds to a spatial scale of $\approx140R_{\rm sch}$\footnote{We note a factor of 2 uncertainty in the assumed BH mass for M\,87, depending
on the BH mass determination method (compare  \citealt{walsh13} with \citealt{gebhardt11}).
In this paper, we use $M_{\rm BH}=6.1\times10^{9}M_{\rm BH}$ for consistency with recent VLBI studies of the M87 jet.}.
At 86\,GHz and with a 50 micro-arcsecond VLBI observing beam, it is therefore possible to image the jet base with a spatial resolution of 7~$R_{\rm sch}$.

The structure and dynamics of the M\,87 jet have been extensively studied in previous VLBI observations.
\cite{junor99} discovered a large apparent jet opening angle of $\sim60^{\circ}$ on sub-pc scales,
which is much wider than the opening angle on the kpc-scale jet ($\sim10^{\circ}$).
This suggests ongoing collimation of the jet flow between sub-pc and pc-scales.
An edge-brightened jet morphology and a faint counter-jet were seen in multi-epoch 
Very Long Baseline Array (VLBA) observations at 15 and 43\,GHz \citep{kovalev07,ly07}.
The study of the jet collimation profile by \cite{asada12} and \cite{hada13} revealed 
a parabolic jet shape up to the Bondi radius, and free conical expansion beyond. 
The variation of the jet expansion profile could be explained through a variation of the external pressure
\citep{komissarov07,lyubarsky09}.

In M\,87 the jet kinematics is complex.
Several distinct, moving emission features suggest a systematic acceleration from 
subluminal motion near the VLBI core ($\lesssim 1c$ at $\lesssim0.1$\,pc projected core separation) 
to superluminal velocities on pc and also on kpc scales (up to $6c$) 
\citep{biretta99,cheung07,kovalev07,walker08,asada14,hada16,mertens16}.
However, subluminal motion is also seen at these larger core distances. 
This has been interpreted as plasma instabilities and moving patterns within the jet \citep{mertens16}.

We note that global VLBI observations at 230~GHz with the Event Horizon Telescope (EHT; e.g. \citealt{doeleman12}) 
measure the size of the VLBI components, which are associated with the jet launching region near the event horizon of the central BH. 
\cite{krichbaum14} reported a tentative size of the VLBI core of $\leq 3.5 R_{\rm sch}$ from 
a 3 component Gaussian model fitting to more recent EHT data. 
However, the still limited uv-coverage of the previous EHT observations 
did not yet allow imaging of M\,87 with high fidelity at this frequency.

With these restrictions in mind, 
Global Millimeter VLBI Array observations (GMVA; e.g. \citealt{marti-vidal12,krichbaum14}) 
at 86 GHz play a crucial role in further tracing the M\,87 jet towards its origin.
With the GMVA, it is possible to directly compare the structure of the 
innermost jet of M87 observed at the aforementioned spatial resolution of $7R_{\rm sch}$ 
with GRMHD simulations of the jet formation region
(e.g. \citealt{mckinney06,tchekhovskoy11,moscibrodzka16}).
GMVA 86\,GHz observations are also crucial to bridge the gap between 
the larger-scale jet morphology seen at observing frequencies $\nu \leq 43$\,GHz and
the event-horizon-scale structures expected to be observed at $\nu \geq 230$\,GHz by future EHT observations.

The GMVA observations of M\,87 since 2004 have shown an edge-brightened core-jet structure tracing the jet up to $\sim 3$\,mas separation from the VLBI core \citep{krichbaum06,krichbaum14,kim16}. 
These early observations have been confirmed and complemented by follow up studies with the VLBA and 
the Green Bank Telescope  (GBT) \citep{hada16} which have a factor of $\sim 2$ lower resolution. 
These data also show a limb-brightened jet structure which allow a quantitative study of the jet collimation profile down to 0.1\,mas core separation.
In the GMVA observations of May 2015 the number of participating telescopes was further increased, 
including the GBT, the IRAM 30m telescope, and the phased Plateau de Bure interferometer.
This led to enhanced uv-coverage and imaging sensitivity
and provided a new 86\,GHz VLBI image of M\,87 which we will present in this paper.

In Sect. \ref{sec:data}, we summarize the observations and data reduction scheme.
We present the main results and the analysis in Sect. \ref{sec:results}.
In Sect. \ref{sec:discussions}, we discuss the physical meaning of our findings in a broader context. 
Sect. \ref{sec:conclusion} summarizes the results.

Throughout the paper we assume a $\Lambda$CDM-cosmology with cosmological constants  
$H_{0}=71$\,km\,/\,s\,/\,Mpc,
$\Omega_{m}=0.27$ and 
$\Omega_{\Lambda}=0.73$ \citep{komatsu11}.
Unless specifically mentioned, the length scales in $R_{\rm sch}$ and parsec(s) 
refer to de-projected distances along the jet axis 
based on the jet viewing angle of $18^{\circ}$ determined from the kinematics of the jet  \citep{mertens16}.

\begin{table*}
\caption{
Summary of the 86\,GHz GMVA observations of M87.
Station abbreviations are:
EB---Effelsberg,
ON---Onsala,
PV---the IRAM 30m telescope at Pico Veleta,
PB---the phased Plateau de Bure interferometer (number of the phased antennas in brackets),
GBT---the Green Bank Telescope, and
VLBA---8 VLBA stations equipped with 3mm receiver (without Hancock and Saint Croix).
The synthetic beam sizes are for natural weighting; angle denotes the position angle of the major axis.
Columns denote:
(1) observing date; 
(2) participating stations;
(3) beam sizes : major axis, minor axis (in $\mu$as) and position angle (in deg);
(4) receiver polarization (L=LCP; R=RCP); 
(5) total bandwidth; 
(6,7) peak and rms noise level in the resulting images.
}             
\label{tab:data}      
\centering          
\begin{tabular}{cccccccc}
\hline\hline 
Date & Stations & Beam & Pol & $\Delta \nu$ & Peak & RMS \\
(yyyy/mm/dd) & & ($b_{\rm min} \times b_{\rm maj}$, $b_{\rm PA}$) & & (MHz) & (mJy/beam) & (mJy/beam) \\
(1)  &  (2) & (3) & (4) & (5) & (6) & (7) \\
\hline
 2004/04/19 & EB, ON, PV, VLBA$^{a}$ & $71 \times 267$, $-$9.14 & L & 128 & 411 & 0.26 \\
 2005/10/15 & EB, ON, PV, VLBA & $61 \times 214$, $-$5.04 & L & 128 & 408 & 0.27 \\
 2009/05/09--10 $^{b}$ & EB, ON, PB(6), VLBA & $72 \times 272$, $-$10.3 & L & 128 & 758 & 0.24 \\
 2014/02/26 & GBT, VLBA & $116 \times 307$, $-$9.0 & L\&R & 512 & 496  & 0.14 \\ 
 2015/05/16 & EB, ON, PV, PB(5), GBT, VLBA & $59 \times 273$, $-$6.88 & L\&R & 512 $^{c}$ & 351 & 0.16 \\
%   \hline
\hline
\end{tabular}
\tablefoot{
(a) Brewster was not available;
(b) Observations in 2009 were conducted over two consecutive days;
(c) PB observed with a reduced bandwidth of 256~MHz (1~Gbps recording rate).
}
\end{table*}
%

 %--------------------------------------------------------------------
\section{GMVA observations and data reduction}\label{sec:data}

M\,87 has been observed by the GMVA during 2002 and 2015 \citep{krichbaum06,krichbaum14,kim16}.
In all observations M\,87 and the calibrator 3C\,273 were observed in full uv-tracks, with up to 15 hrs mutual visibility between the stations.
Here we focus on the data obtained after 2004 which allow reliable imaging of the core and the limb-brightened jet structure.
Between 2004 and 2009 the data were recorded at a bit rate of 512~Mbps in left circular polarization (128\,MHz bandwidth). 
The observations in 2015
were performed at a higher bit rate of 2~Gbps and in dual circular polarization, resulting in the total bandwidth of $2\times256$~MHz.
The GMVA data from observations before 2015 were re-analyzed for this study.
We also included one archival data set from 2014 with VLBA+GBT observations \citep{hada16} in order to enhance the time sampling.
The observational details and resulting imaging parameters are summarized in Tab. \ref{tab:data}.

The GMVA data were correlated with the Mark IV correlator at the Max-Planck-Institut f\"ur Radioastronomie (MPIfR) in Bonn, Germany.
The 2015 data were correlated with the DiFX correlator \citep{deller07} at the MPIfR.
The post processing was performed in AIPS \citep{greisen90} following standard calibration procedures.
The fringe fitting was done in two steps.
First, we removed constant sub-band delays and phase offsets in different IFs using high SNR scans (manual phase-cal).
After the phases were aligned across the frequencies, the data were fringe-fitted using the full bandwidth.
In this global fitting step, residual single and multi-band delays were removed and the fringe rates were determined. 
For reliable fringe detection we applied a cutoff of SNR $\geq$ 5.
The a-priori amplitude calibration was performed using the station based system temperature measurements and gain-elevation curves. From the elevation dependence of the system temperatures an atmospheric opacity correction was derived and applied for each station. After these basic calibration steps, 
the data were averaged over frequency and exported to the Difmap VLBI imaging package \citep{difmap}, 
where the final coherent time averaging (coherence time $\sim10\,s$) and data editing was performed.
For the imaging, we used the CLEAN algorithm \citep{hogbom74} implemented in Difmap. 
An iterative process of phase and amplitude self-calibration was applied until the rms noise level in the map was minimized.
The off-source rms level in each final CLEAN map was estimated using the AIPS task IMEAN, which reads a CLEAN image and fits a Gaussian to the histogram of the pixel values. 
In addition to the thermal noise, we also estimated the systematic uncertainty of the absolute flux to be $\sim 15$\%
based on the station gain corrections derived during the amplitude self-calibration.

\section{Results and Analysis}\label{sec:results}

\subsection{Visibilities and resulting images}\label{subsec:vis_image}

\begin{figure}[t!]
\centering
\includegraphics[width=0.48\textwidth]{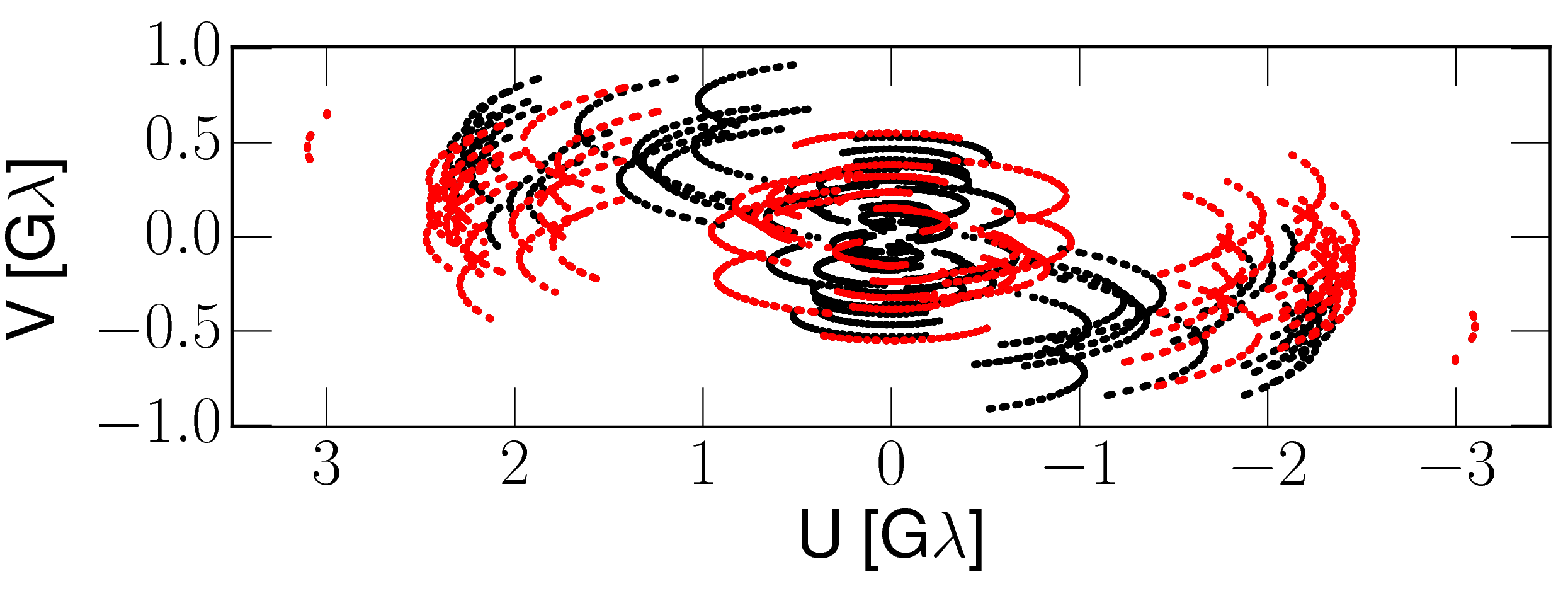}
\caption{
$(u,v)$-coverage for M\,87 from GMVA observations in May 2015.
Only scans whose VLBI fringes have been detected are shown.
Baselines including high-sensitivity stations (PV, PB, and GB) are marked in red.
}
\label{fig:uvcover}
\end{figure}

\begin{figure}[t!]
\centering
\includegraphics[width=0.48\textwidth]{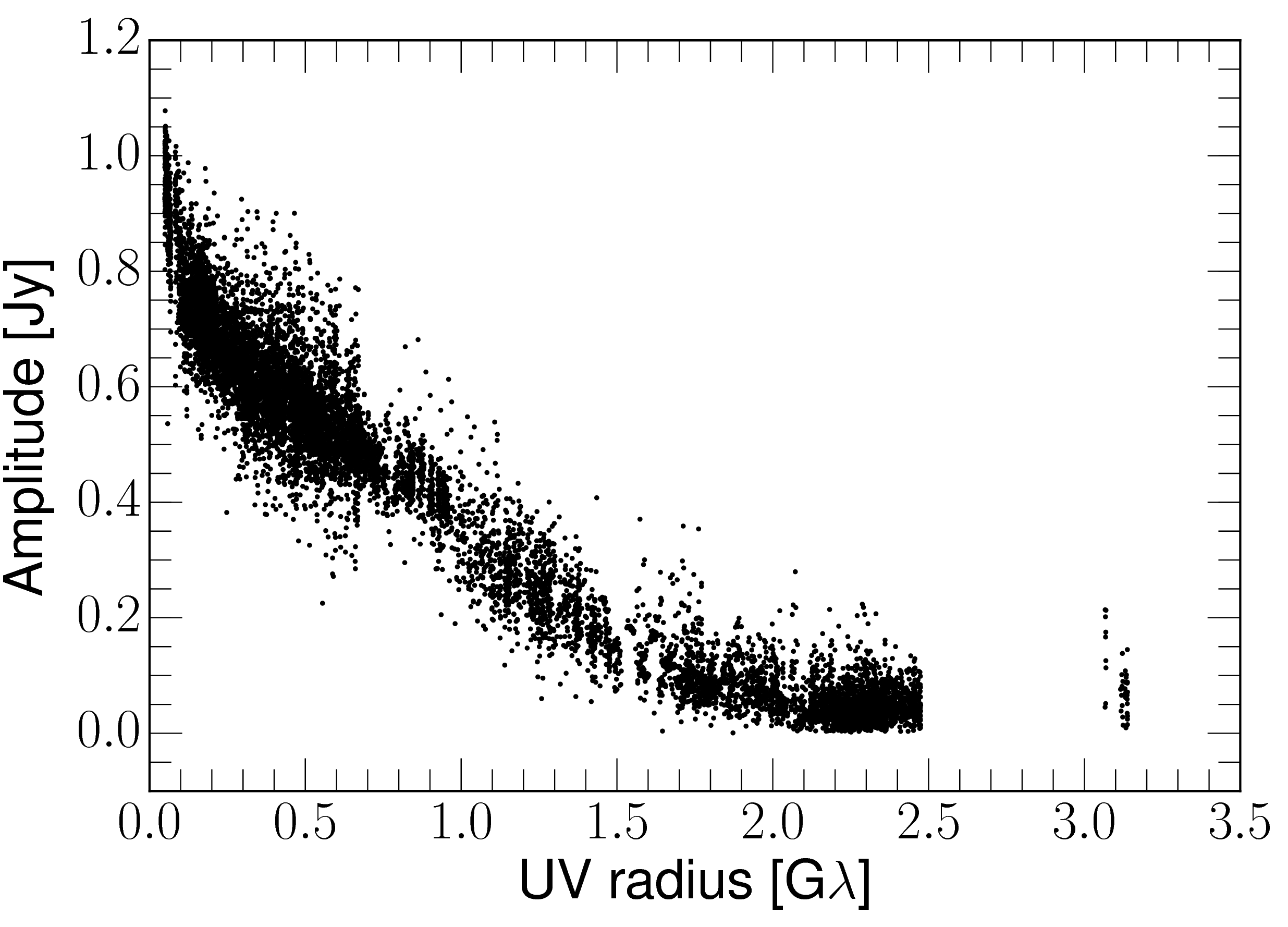}
\caption{
Radial distribution of the visibility amplitude of M\,87 for the GMVA observation at 86\,GHz in May 2015.
For clarity, the data were binned in 30s time intervals.
}
\label{fig:uvplot}
\end{figure}

\begin{figure}[t]
\centering
\includegraphics[width=7.5cm]{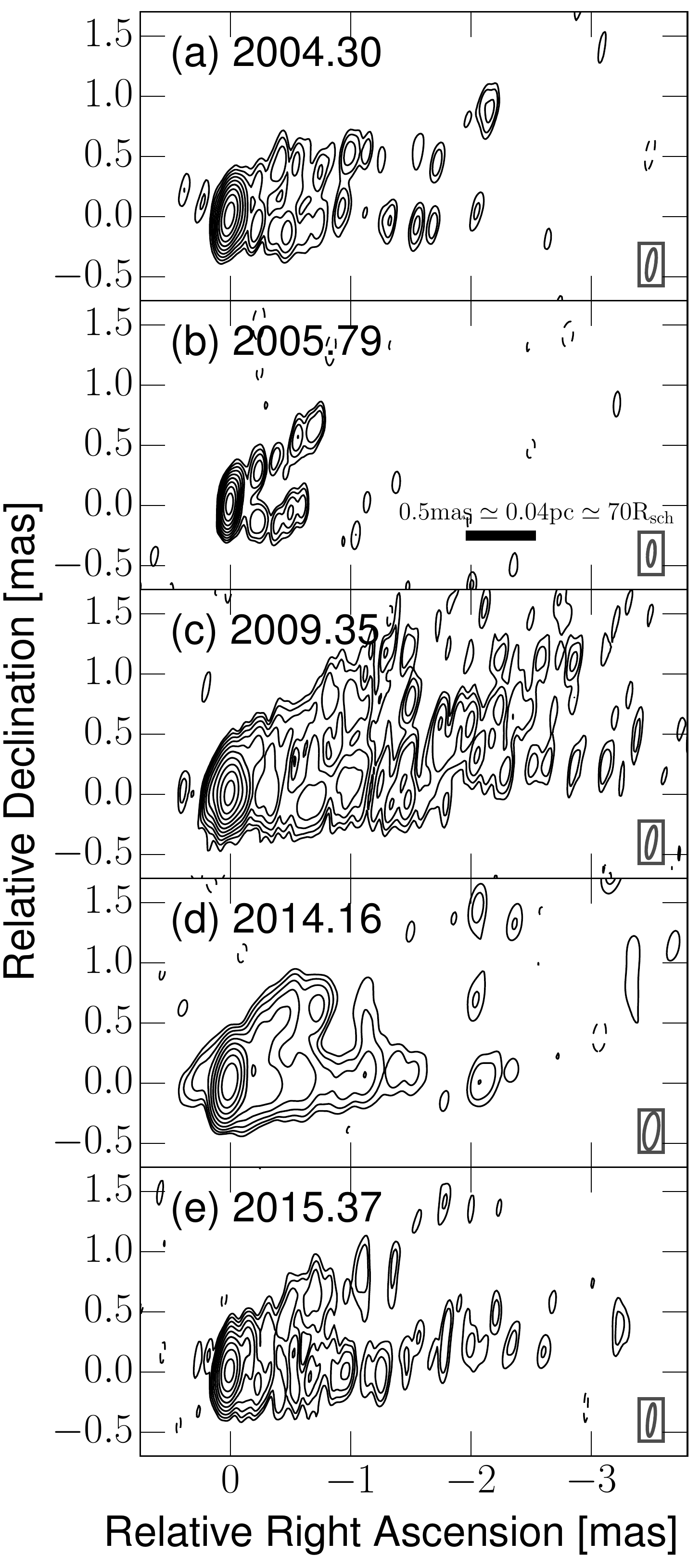}
\caption{
  86 GHz VLBI images of the inner jet in M\,87 obtained from observations
  between 2004 (top) and 2015 (bottom).
  The restoring beam is shown as ellipse at the bottom right corner of each 
  map. The contour levels are $(-1, 1, 2, 4, 8, ...)\times1$~mJy/beam.
        }
\label{fig:grid}
\end{figure}

In Fig. \ref{fig:uvcover} we show the $(u,v)$-coverage of the GMVA observations in 2015.
The inclusion of the large and sensitive telescopes PV, PB, and GBT 
improved the data quality and gave robust fringe detections even on the longest baselines (up to $\sim3$G$\lambda$).
In Fig. \ref{fig:uvplot} we show a representative plot of the radial dependence of the visibility amplitudes versus the $uv$-distance.
The visibilities drop from $\sim1$~Jy at short uv-spacings to $\sim50-100$\,mJy at the longest $uv$-distances 
(e.g., 3G$\lambda$, PV and PdB to MK).
Figure \ref{fig:grid} shows the 86~GHz images of the inner jet region of M87 between 2004 and 2015 (forward in time from top to bottom).
The images show the basic source structure (core-jet morphology with limb-brightening) is similar in all epochs
(though the data and image quality differ between observations).
In addition to the prominent east-west oriented jet, faint emission east of the bright VLBI core is visible. 
This feature may be the counter-jet which is also visible at longer wavelengths
\citep{kovalev07,walker16}. 
The validity of our counter-jet detection at 86\,GHz was tested by applying different phase and amplitude self-calibration schemes
which do not allow removal of this feature from the final image. 
We also point out that fine-scale structure of the jet and the positions of several bright emission features change with time.
We note that our sparse time-sampling does not allow us to robustly cross-identify these bright components across different epochs, which is not unexpected in view of the complex nature of the kinematics in the inner jet \citep{mertens16}. We note that for the two-days long experiment in 2009 
no significant signature of flux variability or motion in the jet was detected over a timescale of two days. 
We therefore combined the two-days long experiment into a single visibility data set and made a single image 
(Figure \ref{fig:grid}c).

\subsection{VLBI-core properties}\label{subsec:core_result}

In order to measure the properties of the VLBI core, %which we in the following associate with the jet base at 86\,GHz (so called VLBI core),
we fitted elliptical Gaussian components to the fully calibrated visibilities using the Modelfit task in the Difmap package.
The model-fit provided us the core flux $S_{\rm core}$, the FWHM size $\psi_{\rm maj, min}$
along the major and minor axes of the ellipse, and
the position angle of the major axis $\rm PA_{\rm core}$ for each epoch.
To estimate uncertainties in the model-fit parameters we followed \cite{schinzel12} in which the authors
accounted for the effects of the finite signal-to-noise and strong side-lobe interference.
The ratio of the peak to the rms noise near the core is $\gtrsim70$ in all epochs.
For this signal-to-noise, the errors of our model-fit parameters are approximately $\sim$15\% and $\sim20$\% for the flux density and the core size, respectively.

The model-fit parameters for the VLBI core are shown in Tab. \ref{tab:core}.
In all cases, the FWHM size of the core is larger than 64\% and 23\% of the beam along its minor and major axis, respectively.
These sizes are larger than the empirical resolution limit of 1/5th of the beam.
If we take the signal-to-noise of $\sim70$ into account, model-fit components whose sizes are larger than $\sim$10\% of the beam 
can be claimed to be spatially resolved \citep{lobanov05,schinzel12}.
Thus we conclude that the core of M\,87 is spatially resolved 
(See also \citealt{baczko16} for a more detailed discussion about 
determination of the resolution limit of a typical GMVA data set).
The flux density of the VLBI core is in the range of $\sim(0.53-0.67)$\,Jy for all observations except 2009, when a significantly higher (factor 2) core flux was seen.
We checked the amplitude calibration using 3C~273, which was observed in alternate VLBI scans in the same experiment and found no evidence for a systematic amplitude mis-calibration.
We therefore regard the elevated flux density as intrinsic to the source. Details regarding this particular epoch will be studied in a future publication.

\subsection{Image stacking}\label{subsec:stack}

Owing to the sparse time sampling of our observations, we focus on an approach 
to average all the images in time (image stacking), in order to increase the image sensitivity and 
study the time-averaged emission, which characterizes the overall shape of the jet launching region.
Such stacking analysis has also been adopted in other studies, e.g.,
in the analysis of the jet structure of AGN at 15\,GHz and other frequencies (e.g. \citealt{fromm13,macdonald15,boccardi16_43g,pushkarev17}).

For the stacking, the individual jet images (Fig. \ref{fig:grid}) were restored with a common beam. The restored maps were aligned by the positions of their intensity peaks. 
The assumption of a stable core position is supported by astrometric VLBA 43 GHz observations of M\,87 over multiple epochs \citep{acciari10}, which revealed a stationary peak position on scales of $6~R_{\rm sch}$.
We performed the stacking procedure with two different beam sizes:
(i) with a larger beam size of $0.1\times0.3$~mas in order to recover the faint jet on $1-4$\,mas core separation, and
(ii) with a smaller beam of 0.051$\times$0.123 mas, in order to reveal the fine scale structure on $<1$\,mas.
The latter beam corresponds to 50\% super-resolution in N-S direction for uniformly weighted data. %the resolution of the uniformly weighted 2015 data.

In the top panel of Fig. \ref{fig:stack} we show the stacked image with the larger beam (Fig. \ref{fig:stack}a) 
and below with the smaller beam (Fig. \ref{fig:stack}b).
The 1$\sigma$ rms level in the higher-resolution stacked image is $\sim0.1$\,mJy/beam, 
making it the deepest and highest resolution view of the M87 jet structure to date.

\subsection{Transverse emission profile and central emission features}\label{subsec:slice}

\begin{figure*}[t]
\centering
\includegraphics[width=0.8\textwidth]{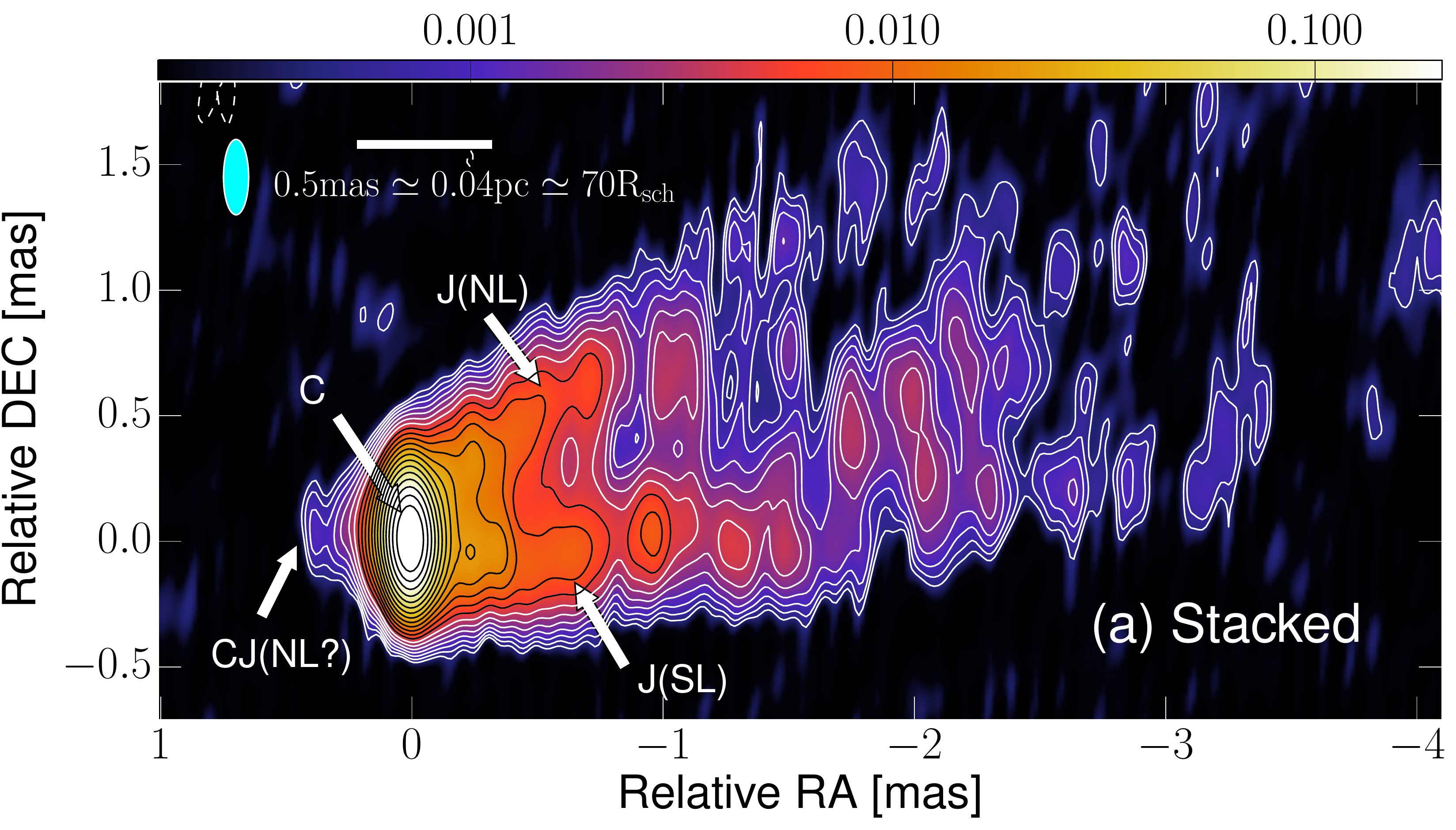}
\includegraphics[height=0.42\textwidth]{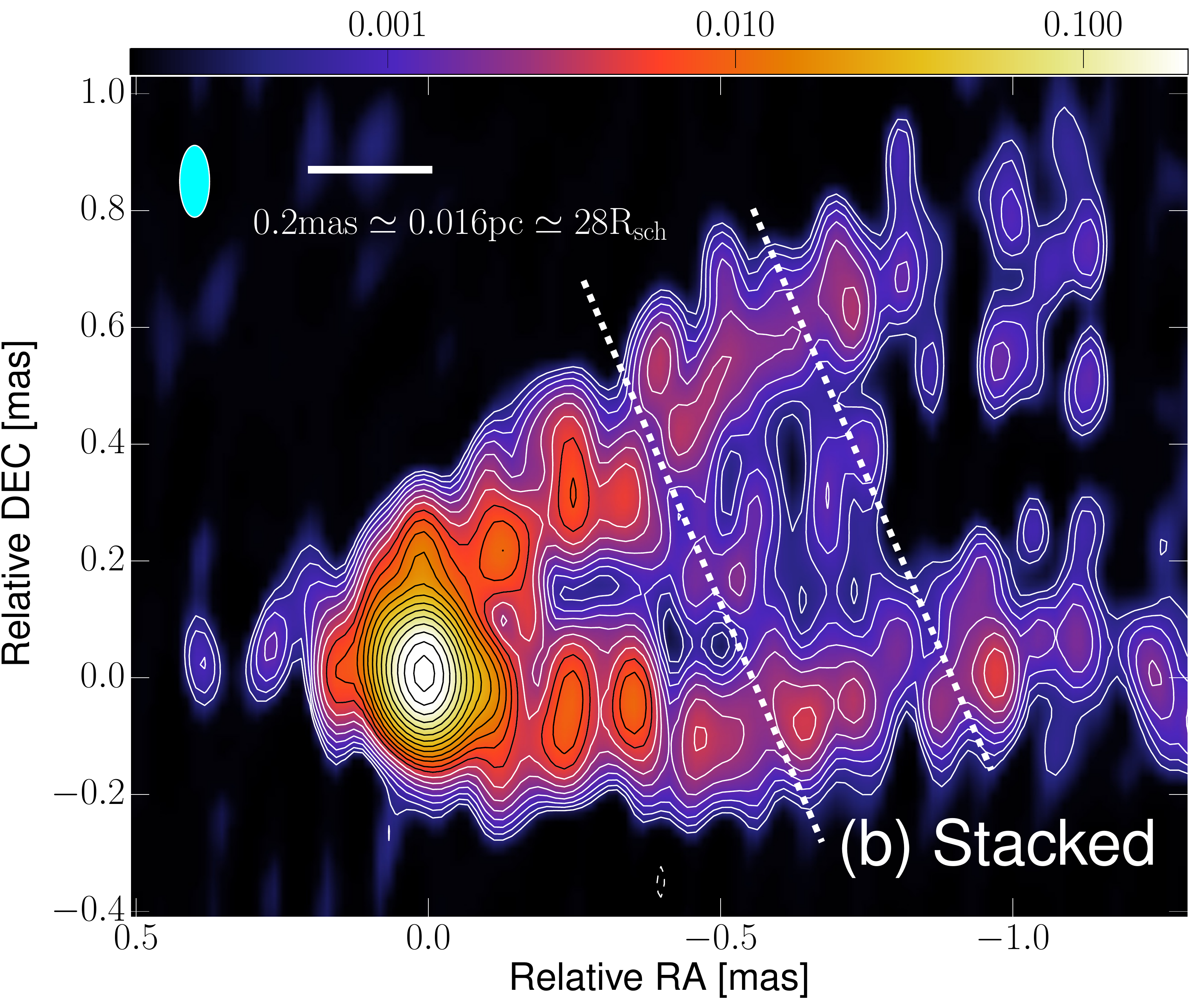}
\includegraphics[height=0.4\textwidth]{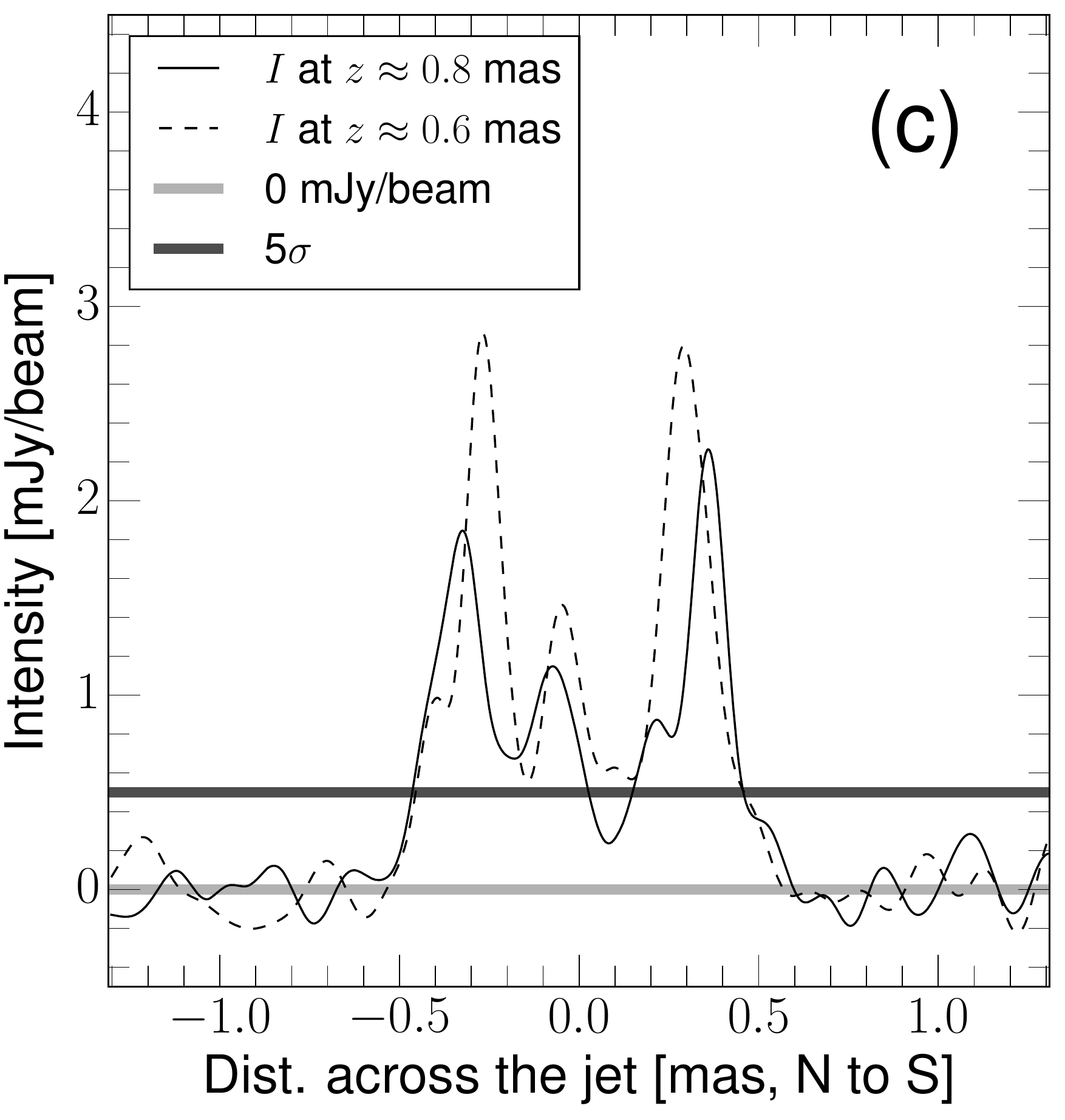}
\caption{
Stacked M\,87 jet images and transverse intensity profiles.
Panel (a) : The image with a restoring beam of $0.3\times0.1$~mas.
The core (C), northern/southern limbs (NL/SL) of the jet (J), and the counter jet (CJ) are indicated by white arrows.
Panel (b) : The same image but restored with a smaller beam of $0.123\times0.051$~mas and zoomed in on the inner region.
The colorbars indicate total intensities in units of Jy/beam.
Contour levels are $(-1, 1, 1.4, 2, 2.8, ...)\times0.47$~mJy/beam.
The white bars indicate projected linear distance scales for M\,87.
The white dashed lines denote the position of the slices in panel (c).
The restoring beams are indicated by the cyan ellipses at the top left corner of each panel.
Panel (c) : 
The transverse jet intensity profiles measured by using the higher resolution image in Fig. \ref{fig:stack}b 
(starting from north to south).
The dark solid/broken lines are the measured intensity at $\sim0.8/0.6$\,mas core distance, respectively.
The light gray line is the zero intensity level and the dark thick gray line indicates the 5$\sigma$ level. 
}
\label{fig:stack}
\end{figure*}

\begin{figure}[t]
\centering
\includegraphics[width=8.5cm]{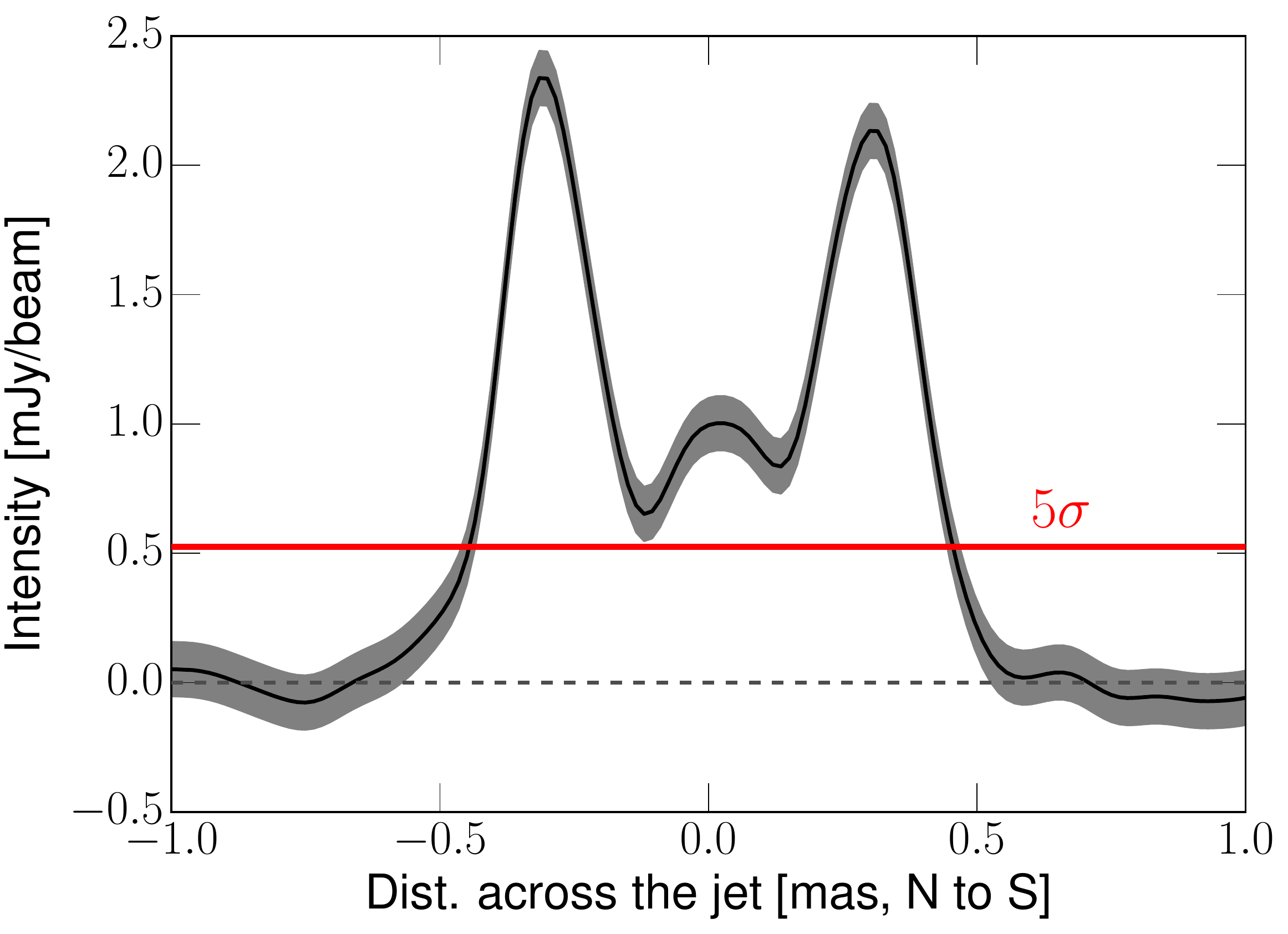}
\caption{
Transverse jet intensity profile obtained by averaging the jet emission over $\sim0.5-0.9$\,mas core distance.
The dark solid line is the measured intensity, 
the shaded region indicates $1\sigma$ image rms noise level, and 
the red line denotes the corresponding $5\sigma$ level.
}
\label{fig:slice_avg}
\end{figure}

We measure the transverse width of the jet with cuts perpendicular to the mean jet axis. 
In Figure \ref{fig:stack}c we highlight the transverse intensity profiles in two cuts 
made at $\sim0.6$\,mas and $\sim0.8$\,mas core separations.
The transverse jet profiles appear limb-brightened, with an additional signature of a fainter central emission component
that is visible above the 5$\sigma$ image noise level.
In order to characterize the brightness of this central component, 
we calculated the center-to-limb brightness ratio $\rho_{\rm CL}$.
We integrated the fluxes in the central lane and the two limbs over $(0.45-0.95)$~mas core distance and 
we obtained $\sim38$~mJy and $\sim15$~mJy for each limb and the center, respectively.
Accordingly, we find $\rho_{\rm CL}\sim0.4\pm0.1$. 
For this calculation we assumed 15\% for the flux density uncertainty for the limb but 30\% for the fainter central lane.
As an illustration, we show the averaged transverse intensity profile in Fig. \ref{fig:slice_avg}.

\begin{table*}
\caption{
Properties of the VLBI core estimated by elliptical Gaussian model fitting.
}
\label{tab:core}      
\centering          
\begin{tabular}{c c c c c c c c c}
\hline\hline 
Epoch & 
$S_{\rm core}$ & 
$\psi_{\rm min}$ & 
$\psi_{\rm maj}$ & 
\multicolumn{2}{c}{$\sqrt{\psi_{\rm maj}\times\psi_{\rm min}}$} &
$\rm PA_{\rm core}$ & 
$T_{\rm B}\times\delta$ \\
(yyyy/mm/dd) & (Jy) & ($\mu$as) & ($\mu$as) & ($\mu$as) & ($R_{\rm sch}$) & (deg, N$\rightarrow$E) & ($\times10^{10}K$) \\
(1)          &  (2)  &   (3)     &  (4)      & (5)  & (6)                   & (7)  & (8) \\
\hline
2004/04/19 & 0.58$\pm$0.09 & 60$\pm$12 &  63$\pm$13 & $62\pm9$ & $8.7\pm1.2$ & $29.4$ &  2.5$\pm$0.8 \\
2005/10/15 & 0.53$\pm$0.08 & 38$\pm$ 8 &  88$\pm$18 & $58\pm8$ & $8.1\pm1.1$ & $-11.0$ & 2.6$\pm$0.8 \\
2009/05/09 & 1.39$\pm$0.20 & 75$\pm$15 & 127$\pm$30 & $102\pm15$ & $13.7\pm1.9$ & $15.6$ & 2.4$\pm$0.8 \\
2014/02/26 & 0.60$\pm$0.09 & 75$\pm$15 &  80$\pm$16 & $77\pm11$  & $10.8\pm1.5$ & $18.6$ & 1.7$\pm$0.5 \\ 
2015/05/16 & 0.67$\pm$0.10 & 87$\pm$18 &  91$\pm$18 & $89\pm12$  & $12.5\pm1.8$ & $48.0$ & 1.4$\pm$0.5 \\
\hline
\end{tabular}
\tablefoot{
Each columns show 
(1) the observing epoch (in year/month/day);
(2) the core flux (in Jy);
(3) and (4) the FWHM core size along the minor and major axis (in $\mu$as);
(5) and (6) the geometrical mean of the core size (in $\mu$as and $R_{\rm sch}$, respectively);
(7) the position angle of the elliptical core (in deg) and 
(8) the Doppler-boosted apparent brightness temperature (in $10^{10}$\,K).
}
\end{table*}

\subsection{Ridge line analysis}\label{sec:ridge_analysis}

In order to analyze the jet base structure using the ridge lines, the following procedure was performed.
We took the high-resolution jet image in Fig. \ref{fig:stack}.b and rotated the image by $21^{\circ}$ in the clockwise direction
assuming that the overall jet position angle $PA_{\rm c}$ is $-69^{\circ}$ with respect to north.
At each core distance $d$, we made slices transverse to the jet and fitted two Gaussians to the two humps in the transverse intensity profile.
We were often forced to fit another Gaussian along the jet center axis to improve the fit quality.
We began this fitting procedure at a core distance $d\sim1$~mas and continued fitting the transverse profile 
down to $d\sim0.06$\,mas.
We followed \cite{mertens16} in computing the uncertainties in the ridge line analysis.
The diameter of the jet $W(d)$ has been determined by
the distance between the positions of the Gaussian peaks at each $d$.
Similarly, we also derived the apparent jet opening angle $\phi_{\rm app}(d)$ 
by computing the angle subtended by the two Gaussian peaks.
The intrinsic opening angle $\phi_{\rm int}$ is then given by 
$\phi_{\rm int}=2\arctan(\sin\theta\times\tan(\phi_{\rm app}/2))$
\citep{pushkarev17}
under the assumption that the extended jet is azimuthally symmetric. 

\begin{figure}[t]
\centering
\includegraphics[width=8.5cm]{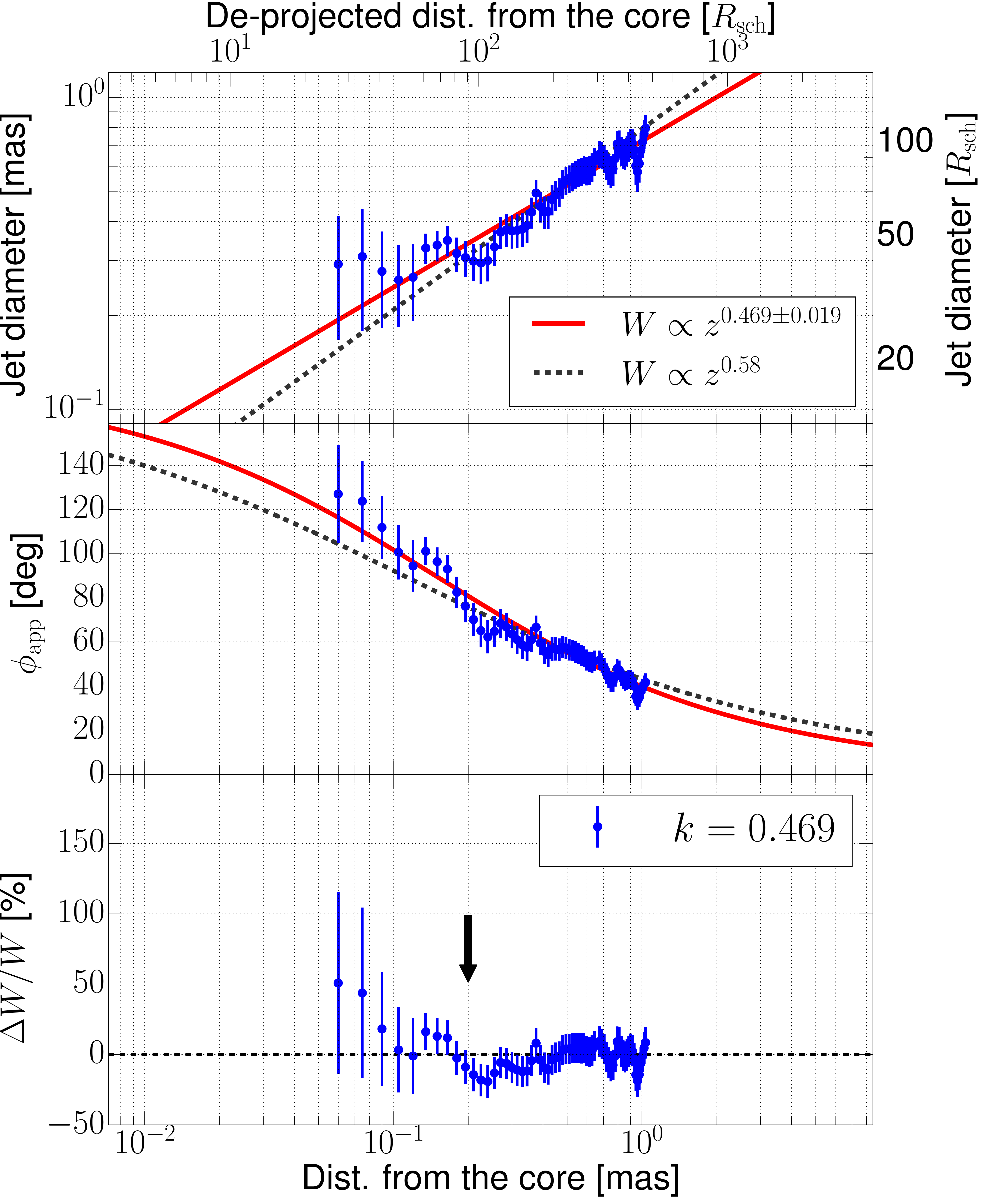}
\caption{
The M\,87 jet base collimation profile plotted versus the core separation $d$.
\textit{Top} : the jet width and the power-law fit.
The blue data points are the measurements, 
the red solid line is the best power-law fit to the data, and
the grey broken line is a fit with a fixed slope of $k=0.58$.
\textit{Middle} : The same but for the apparent opening angle.
\textit{Bottom} : the fractional difference between the observed diameter and the power-law model 
with a corresponding slope, $k\approx0.47$.
The arrow indicates a core distance where the fractional difference starts to grow significantly.
}
\label{fig:collimation}
\end{figure}

Figure \ref{fig:collimation} shows the results of the ridge line analysis.
The jet diameter at $d<1$\,mas apparently increases with core separation.
The overall jet diameter at $d>0.3$\,mas is $W\sim0.3-0.7$~mas.
At smaller core separations ($d<0.2~$mas, 90$R_{\rm sch}$ projected), 
the collimation profile slightly flattens and the jet diameter does not sharply decrease.
At the jet base ($d=0.06$~mas), the jet width is $0.29\pm0.09$\,mas ($40.8\pm12.3~R_{\rm sch}$)
with an apparent opening angle of $\phi_{\rm app}=127^{\circ} \pm 22^{\circ}$.
Depending on the jet viewing angle, the intrinsic jet opening angle is $\phi_{\rm int}=63.6^{\circ}\pm25.0^{\circ}$ ($90^{\circ}\pm28^{\circ}$)
for a jet viewing angle $\theta=18^{\circ}$ (30$^{\circ}$), respectively.

\begin{figure}[t]
\centering
\includegraphics[width=0.45\textwidth,trim={10pt 50pt 10pt 50pt},clip]{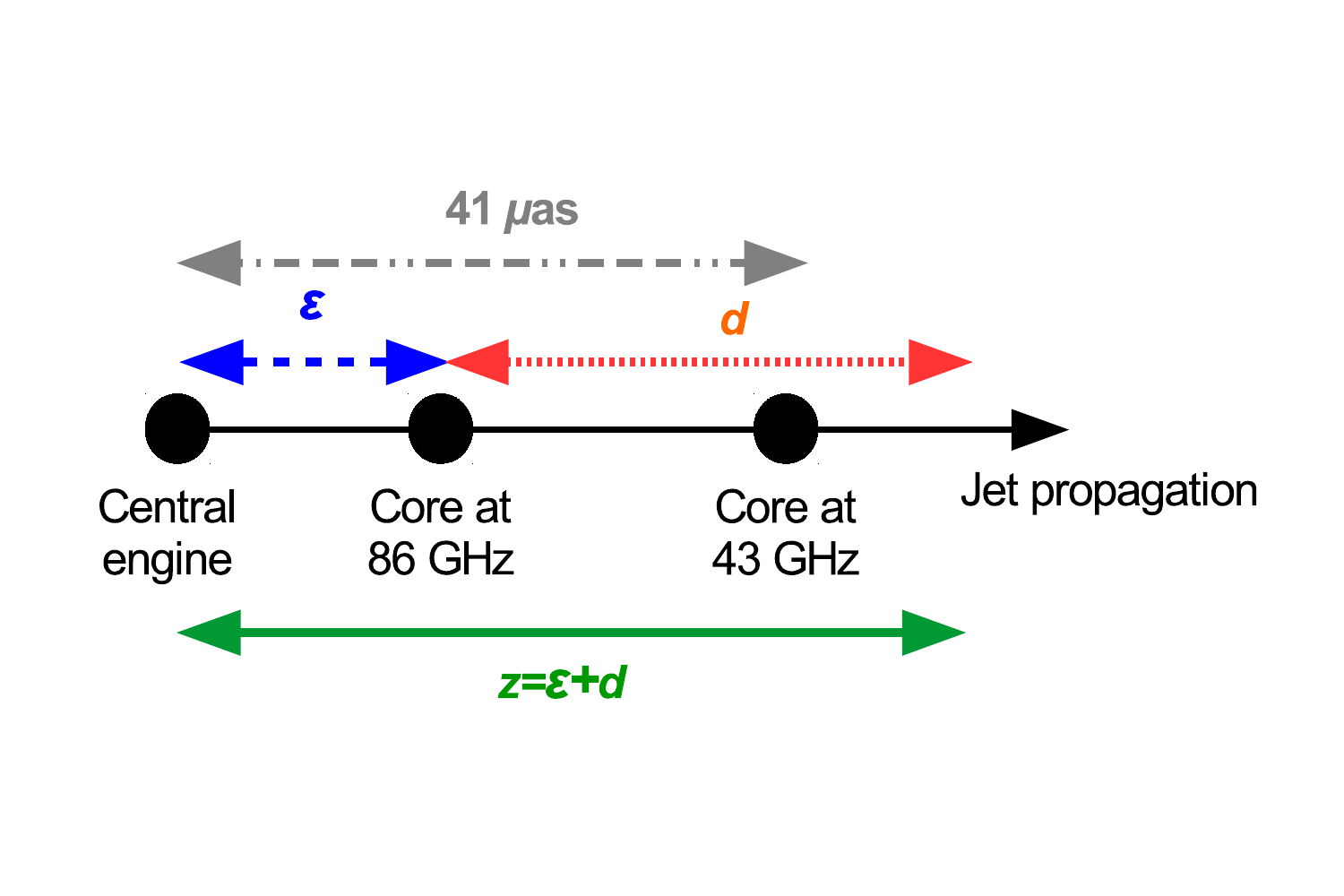}
\caption{
Illustration of the central engine and the 86~GHz core geometry described in Sect. \ref{sec:ridge_analysis}.
}
\label{fig:illust_geometry}
\end{figure}

Asymptotic structure in the measured jet width $W$ versus the distance from the central engine $z$ 
was fit with a power-law model $W(z) \propto z^{k}$ where $k$ is a dimensionless index
which parameterizes the jet expansion and acceleration within theoretical models 
(e.g. \citealt{komissarov07,lyubarsky09}).
It is important to note that the core separation $d$ is not necessarily the same as the
the distance from the central engine $z$ because of the jet opacity (e.g. \citealt{lobanov98}).
Hence, we associate the core separation $d$ to the distance from the central engine by $z=\epsilon+d$
where $\epsilon$ is the unknown offset between the BH and the 86~GHz core (see Fig. \ref{fig:illust_geometry}).
We adopted $\epsilon\leq41~\mu$as based on the results of \cite{hada11},
where the authors performed astrometric VLBA observations toward M\,87 at $2.3-43.2$\,GHz
and estimated the distance between the intensity peak and the jet apex at 43\,GHz.
Then we obtained $W$ as function of $z$ and the power-law model was fit to $W(z)$.

We find a jet expansion rate of $k=0.469\pm0.019$ when we ignore the core-shift at 86\,GHz (i.e., $\epsilon=0$).
With a non-zero core-shift correction ($\epsilon\neq0$), 
we find $k\sim0.47-0.51$ with a mean value of $k=0.498\pm0.025$
(the error represents uncertainties in both the core position and the statistical fitting).
Both fits have a reduced chi square of $\chi^{2}_{\rm red}\sim0.51$.
This is in agreement with previous values of $k=0.56-0.60$ \citep{asada12,hada13,mertens16} within 3$\sigma$ uncertainty levels.
To demonstrate the goodness of fit,
we calculated the fractional difference between the model and the data, 
$\Delta W/W=(W_{\rm obs}/W_{\rm fit} - 1)$ where $W_{\rm obs}$ and $W_{\rm fit}$ are the 
observed and model jet widths, respectively.
The results are displayed in the bottom panel of Fig. \ref{fig:collimation}.
It can be seen that the fractional difference is nearly zero, but starts to grow within $d\leq0.2$~mas.

\subsection{The jet to counter-jet ratio}\label{subsec:cj}

We measured the integrated jet and counter-jet flux density at core separations of
$0.2-0.5$\,mas (projected distances of $\sim 28-140~R_{\rm sch}$) 
using the jet image in Fig. \ref{fig:stack}a.
For the main jet, we placed a large box covering all three elements of the approaching jet 
(i.e., the northern limb, the southern limb, and the central emission).
Another box of the same size was placed in the counter-jet region.
The measured flux density of the jet and the counter jet were $\sim$95~mJy and $\sim3.5$~mJy, respectively.
Accordingly, we obtained the jet to counter-jet ratio $R = 27.1\pm9.1$
(assuming 15\% and 30\% of flux uncertainties for the approaching and the receding jets, respectively).
We note observational evidence for limb-brightening in the counter-jet at 43\,GHz \citep{walker16}. 
This is not seen in our data (possibly) due to sensitivity limitations and the higher frequency.
Therefore, our counter-jet flux could be under-estimated.
In order to correct for this, we have measured the integrated flux of only the southern limb of the approaching jet 
at the same core distances assuming that we see only the northern edge of the limb-brightened counter-jet.
The integrated flux of the southern-limb of the approaching jet was reduced to $\sim60$\,mJy.
This lowers the jet to counter-jet ratio to $R=17\pm6$.

\begin{figure}[t]
\centering
\includegraphics[width=0.40\textwidth]{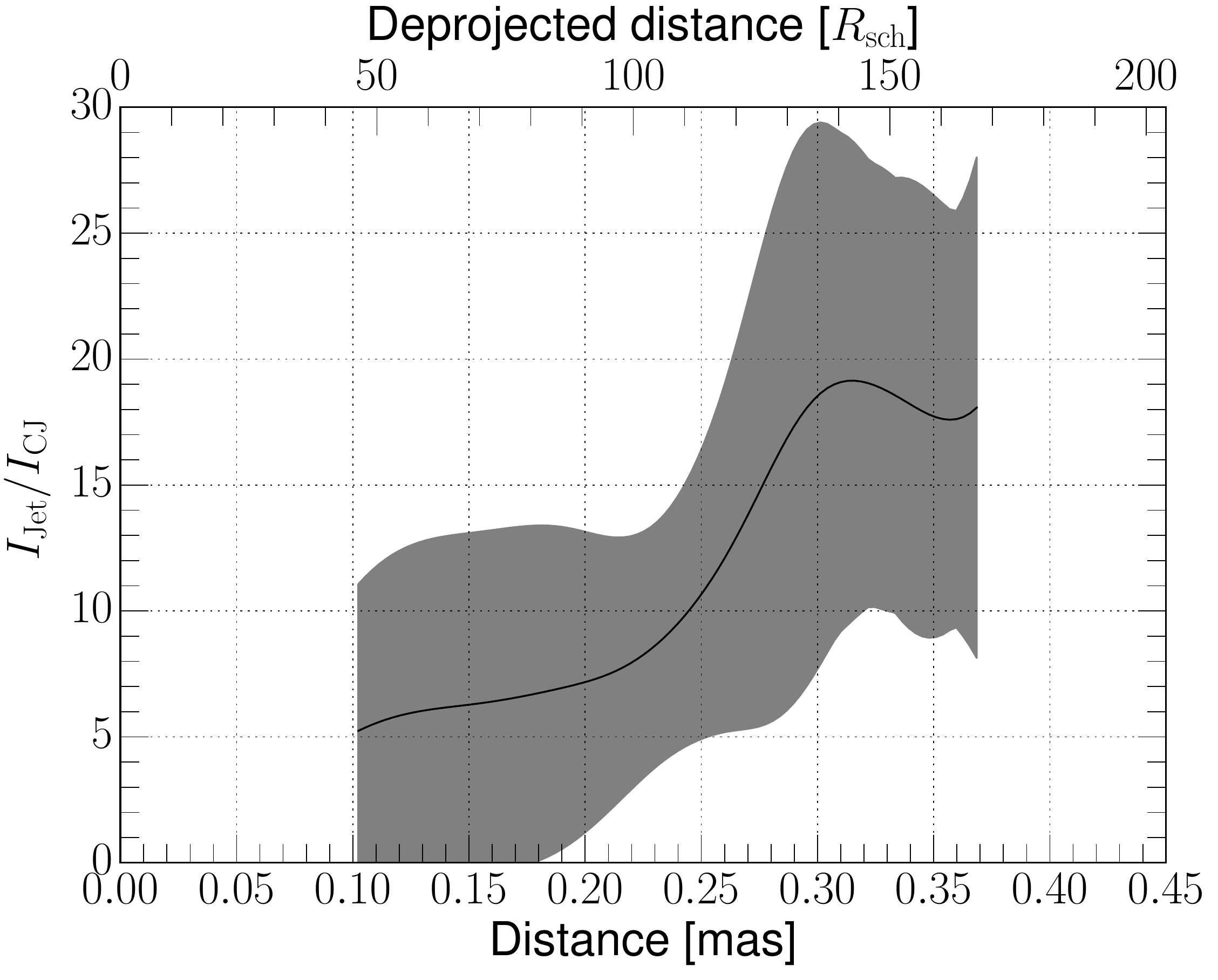}
\caption{
The jet to counter-jet intensity ratio ($R$) measured as described in the text.
The solid line shows the measured values and 
the shaded region indicates the uncertainties.
A viewing angle of $\theta=18^{\circ}$ was used to
calculate the deprojected distances.
}
\label{fig:jet_cj_ratio}
\end{figure}

We also measured the jet to counter-jet ratio variation in the longitudinal direction using individual pixel values.
For this, we cut the jet longitudinally through the counter-jet, the core, and the southern limb of the approaching jet. 
We took the jet image in Fig. \ref{fig:stack}a to obtain a smoother jet intensity gradient.
For a more reliable measurement, we used only pixel values over the 7$\sigma$ level (0.77~mJy/beam).
When calculating the jet to counter-jet ratio as function of the distance from the central engine, 
using the same jet apex to core offset $\epsilon$ may affect our measurements 
(see Fig. \ref{fig:illust_geometry} for illustration).
Therefore, we adopted the same $\epsilon\leq41\,\mu$as in a direction east of the VLBI core 
(highlighted in gray in Fig. \ref{fig:illust_geometry}; see Sect. \ref{sec:ridge_analysis} for details)
and calculated the brightness ratio at the corresponding jet distance.
We accounted for thermal noise, systematic amplitude error, and the dispersion in the $R$ profile at each distance $z$ 
due to the uncertainty in the central engine position.
We further excluded $R$ values measured within 0.1\,mas of the central engine because of the relatively large beam size 
in the image presented in Fig. \ref{fig:stack}a.
The result is shown in Fig. \ref{fig:jet_cj_ratio}. 
The $R$ measured within $\sim0.2$\,mas from the central engine is largely affected by 
the positional uncertainty of the 86\,GHz VLBI core, $\epsilon$, causing $R$ to vary by a factor $\sim5$.
Nevertheless, we find that a single constant value of the jet-to-counter jet ratio is not suitable 
in describing the relatively large variation of $R$ ($R=1-10$ near 0.1~mas and $R=10-25$ at larger distances).

\section{Discussion}\label{sec:discussions}

\subsection{Physical conditions in the VLBI core region}\label{subsec:core_discuss}

The VLBI core brightness temperature parameterizes the physical conditions 
within compact energetic jet components (e.g., \citealt{kovalev05,lee16}).
We calculate the intrinsic brightness temperature of the core in the source rest frame, $T_{\rm B}$, by
\begin{equation}
 T_{\rm B}=1.22\times10^{12}\frac{S_{\rm core}}{ \nu^{2} \psi_{\rm maj}\psi_{\rm min} }\frac{(1+z)}{\delta} \quad \rm K
 \label{eq:tb}
\end{equation}
\citep{lee16} where 
$\nu$ is the observing frequency in GHz, 
$S_{\rm core}$ is in Jy, 
$\psi_{\rm maj}$ and $\psi_{\rm min}$ are in mas, 
$z=0.00436$ is the redshift of M\,87 \citep{smith00}, and
$\delta$ is the Doppler factor.
The observed apparent brightness temperature, $T_{\rm B, app}=T_{\rm B}\times\delta$, is shown in Tab. \ref{tab:core}.
We find that the $T_{\rm B, app}$ is generally quite low.
Remarkably, the $T_{\rm B, app}$ is always lower than $T_{\rm eq}\sim(5\times10^{10})~K$, 
which is often referred to as the brightness temperature of a plasma in which 
the energy density of the magnetic field $u_{\rm B}$ is equal to that of the particles 
$u_{\rm p}$ (i.e., the ``equipartition brightness temperature''; \citealt{readhead94}).
The intrinsic brightness temperature will be even lower if Doppler boosting is accounted for.

In order to examine whether the equipartition brightness temperature for the non-thermal electrons in the jet of M\,87
is higher or lower than $5\times10^{10}$~K,
we explicitly calculate the equipartition brightness temperature in the following manner:
We follow the analysis presented by \cite{singal09}.
Assuming equal energy density for particles $u_{p}$ and magnetic field $u_{B}$ in the jet frame (i.e. without Doppler boosting),
we can rewrite Eq. 3 of \cite{singal09} to express $T_{\rm B, eq}$ in the source frame by
\begin{equation}
 T_{\rm B, eq} = t(\alpha)10^{11}
 \left[ \left(\frac{s}{\rm pc}\right)
 \left(\frac{\nu_{\rm turn}}{\rm GHz}\right)^{1.5+\alpha}
 \right]^{1/8} 
 \quad {\rm K}
 \label{eq:singal}
\end{equation}
where 
$\alpha$ is the spectral index ($S\propto\nu^{+\alpha}$; note that the sign convention is different from \citealt{singal09}),
$t(\alpha)$ is a numerical function of the spectral index tabulated in \cite{singal09} 
($t(\alpha) \sim 0.3-0.7$ for $\alpha\sim-0.3$ to $-1.5$),
$s$ is the linear size of the emitting region, 
and $\nu_{\rm turn}$ is the synchrotron turn-over frequency.
In the observer's rest frame, $T_{\rm B, eq}$ is boosted by the Doppler factor $\delta$ \citep{readhead94}.
For $s$ we adopt the mean FWHM size of the VLBI core
\footnote{
\cite{marscher83} suggested a correction factor of 1.8 for the Gaussian to spherical size conversion. 
This correction changes the $T_{\rm B, eq}$ by a small factor of $1.8^{1/8}\sim1.08$. Thus we ignore this size correction.
}, $~77~\mu$as ($\sim11R_{\rm sch}\sim0.006~$pc).
The exact values of $\nu_{\rm turn}$ and $\alpha$ on $\lesssim11R_{\rm sch}$ scale are not well determined 
due to the lack of simultaneous multi-frequency VLBI observations at high resolution. 
Nevertheless, we compare the time averaged core flux at 86\,GHz ($\sim0.5$~Jy) and at 230\,GHz ($\sim1$~Jy; \citealt{doeleman12,akiyama15})
and find an indication of an inverted spectrum between 86\,GHz and 230\,GHz. 
Also, \cite{broderick09} suggested a turnover frequency $\nu_{\rm turn} \sim 230~$GHz and 
a spectral index $\alpha\sim-1.0$ using other available total flux density measurements.  
Hence we adopt $\nu_{\rm turn}=230$\,GHz and $\alpha=-1.0$ in our analysis.
In addition, we note that Eq. \ref{eq:singal} is insensitive to the exact value of $\nu_{\rm turn}$ and $\alpha$ due to the 1/8 power dependence.
The Doppler factor is given by $\delta=1/\Gamma(1-\beta\cos\theta)$ where 
$\Gamma$ is the bulk Lorentz factor, $\beta$ is the intrinsic jet speed, and $\theta$ is the jet viewing angle.
We adopt the apparent jet speed of $\sim0.5c$ and the viewing angle of $\theta=18^{\circ}$ \citep{hada16,mertens16},
which gives $\delta\sim2$.
Computing Eq. \ref{eq:singal}, we obtain a value of $T_{\rm B, eq}$ in the observer's frame of $\approx7.7\times10^{10}~K$.
For consistency, we also calculate $T_{\rm B, eq}$ using Eq. 4b of \cite{readhead94}, 
which yields $T_{\rm B, eq}\approx3.5\times10^{11}~K$ in the same frame (which is within a factor of 2 of our current value).

Accordingly, the equipartition brightness temperature is larger than the observed values by nearly an order of magnitude.
It is worth noting that the broadband spectral energy decomposition of the core of M\,87 shows 
a large dominance of the jet emission.
In particular, emission from non-thermal electrons in the jet dominate
over the emission from thermal (and non-thermal) electrons in the surrounding accretion disk \citep{broderick09,prieto16}.
We also recall that the observed brightness temperature is still higher than the 
effective temperature of an electron $m_{e}c^{2}/k_{B}\sim6\times10^{9}~$K.
Therefore, the core brightness temperature is a good representation of 
the microscopic energy of the non-thermal electrons in the jet.
Hence, we conclude that in the VLBI core at 86 GHz and on spatial scales of  $\sim11R_{\rm sch}$, 
the magnetic field energy density in the jet is larger than that of the non-thermal particles.

The corresponding magnetic field strength $B$ in the jet can be estimated by 
$B=B_{\rm eq}(T_{\rm B, eq}/T_{\rm B})^{2}$ \citep{readhead94}
where $B_{\rm eq}$ is the equipartition magnetic field strength.
In Tab. \ref{tab:Bfield} we list $T_{\rm B, eq}/T_{\rm B}$ 
and $B/B_{\rm eq}$, adopting an equipartition brightness temperature of
$T_{\rm B, eq}=2\times10^{11}~K$. 
If the characteristic equipartition magnetic field strength in the VLBI core region of M\,87 would match 
the estimates for other AGN on larger scales (of the order of $\sim1$~G, \citealt{pushkarev12_cs}), 
the true magnetic field strength could lie in the range of 61-210~G
at the 86\,GHz jet base and even higher when approaching closer to the central engine. 
This agrees with an independent estimate from \cite{kino15}, who also obtained $B\sim100$~G.
We note that such a strong magnetic field seems to be present in some other AGN-jet systems 
\citep{zamaninasab14,marti-vidal15,baczko16}.
Following \cite{readhead94}, the energy density ratio $u_{\rm p}/u_{\rm B}$ is obtained by
\begin{equation}
 \frac{u_{\rm p}}{u_{\rm B}} = \left(\frac{T_{\rm B, eq}}{T_{\rm B}}\right)^{-17/2}~.
 \label{eq:readhead_upub}
\end{equation}
Using $T_{\rm B, eq}/T_{\rm B}$ tabulated in Tab. \ref{tab:core},
we find $-6\lesssim\log(u_{\rm p}/u_{\rm B})\lesssim-4$. 
Thus, the inner jet of M\,87 appears to be more magnetically dominated.

We note that the above calculations assume a negligible energy contribution from thermal particles
in the accretion flow, which may be entrained in the sheath of the jet (e.g., \citealt{moscibrodzka16}).
The brightness temperature of M\,87 is also quite low compared to typical values found in other sources 
at this frequency ($T_{\rm B,app}\sim10^{11}~$K; \citealt{lee16,nair18}),
perhaps indicating that the non-thermal particles should be at least mildly relativistic.
In this situation, the non-thermal particles in the low energy tail of the particle density distribution
cannot be easily distinguished from those in the thermal particle density distribution.
For instance, we recall that 
near the 86\,GHz the VLBI core the jet plasma is still opaque to synchrotron radiation
(see also \citealt{kim18}).
The self-absorbed synchrotron radiation may heat the low energy particles, 
which will in turn modify the spectral shape of the particle energy distribution at lower energies \citep{ghisellini88}.

\subsubsection{Impact of the thermal particles}

We can make an estimate for the impact of the thermal particles based on a couple of assumptions.
First, 
the number density of thermal particles, $n_{\rm th}$, in the hot accretion flow of M\,87 would be 
two orders of magnitude larger than that of non-thermal particles, $n_{\rm nth}$, 
in order to reproduce the observed source spectrum (e.g., \citealt{broderick09}).
Such a ratio of number densities would remain the same in the jet
if all the particles in the jet are supplied only by the accretion flow.
We note that various particle acceleration mechanisms may operate close to the jet base 
and could make the number density of non-thermal particles larger than what we assume here (e.g., \citealt{mckinney06,pu17}).
Second, the temperature, $T_{i}$, of energetic ions in the hot accretion flow
is nearly virial and can be described by $T_{i}\sim10^{12}\times(z/R_{\rm sch})^{-1}$~K, 
where $z$ is the distance from the central BH \citep{yuan14}.
The electron temperature in such a model is generally much lower than the ions, 
and thus electrons would not significantly contribute to the internal energy of the gas.
If a part of the accreting matter is entrained within the jet at the location of the 86\,GHz core 
and accounts for the internal energy of the thermal particles in the jet,
the energy of the thermal particles in the jet, $u_{\rm th}$, is approximately
$u_{\rm th}\sim n_{\rm th}k_{\rm B}T_{i}$,
where $k_{B}$ is the Boltzmann constant.
The 86\,GHz core is presumably located at $z\sim10R_{\rm sch}$ from the BH (e.g., \citealt{hada11,hada16})
and the corresponding temperature would be $T_{i}\sim10^{11}$~K.
For the non-thermal particles in the jet, 
the energy density $u_{\rm nth}$ can be estimated by 
$u_{\rm nth}\sim n_{\rm nth}\gamma m_{e}c^{2} \sim 6\times10^{9}  \gamma n_{\rm nth}k_{\rm B}$ 
where $\gamma$ is a characteristic particle Lorentz factor.
For mildly relativistic non-thermal particles, we can presume $\gamma$ is order of unity.
Therefore, the energy density ratio between the thermal and non-thermal particles is
$u_{\rm th}/u_{\rm nth} \approx (10^{11}/6\times10^{9})(n_{\rm th}/n_{\rm nth})\sim10^{3}$.
According to our assumption, the jet particle energy density may be dominated by thermal particles.
However, this still does not change the larger dominance of the magnetic field energy
in the total energy budget 
($-3\lesssim \log(u_{\rm p}/u_{\rm B})\lesssim -1$ with $u_{\rm p} \rightarrow 10^{3}u_{\rm p}$).
This implies that the magnetic energy dominates the jet on this spatial scale.

Another important implication of the observed low brightness temperature at 86\,GHz is that 
it is lower than the inverse-Compton limit $T_{\rm B}$ ($\sim10^{12}$\,K; \citealt{kellermann69}), 
which limits the production of high-energy photons.  
Thus, the inverse-Compton scenario disfavors the compact 86~GHz VLBI core region ($\sim11R_{\rm sch}$) as 
the dominant source of gamma-ray or TeV photons. 
Interestingly, the timing analysis of TeV and $\gamma$-ray events associated with the M\,87 nuclear region constrain
the high-energy photon production site to be $\sim20-50R_{\rm sch}$ \citep{acciari10,abramowski12,hada14}, 
which is $2-5$ times larger than the size of the 86~GHz VLBI core.

\begin{table}
\caption{
The brightness temperature ratio $T_{\rm B, eq}/T_{\rm B}$ with $T_{\rm B, eq}=2\times10^{11}$\,K and $\delta\approx2$ 
and the magnetic field strength in units of $B_{\rm eq}$.
}             
\label{tab:Bfield}      
\centering          
\begin{tabular}{ccc}
\hline\hline 
Epoch & $T_{\rm B, eq}/T_{\rm B}$ & $B/B_{\rm eq}$ ($\times10$) \\
(1)  & (2)   & (3) \\
\hline
2004/04/19 & 8$\pm$3 &  6.4$\pm$4.1 \\
2005/10/15 & 8$\pm$2 &  6.1$\pm$3.9 \\
2009/05/09 & 8$\pm$3 &  6.9$\pm$4.4 \\
2014/02/26 & 12$\pm$4 & 15$\pm$9 \\
2015/05/16 & 14$\pm$5 & 21$\pm$13 \\
\hline
\end{tabular}
\tablefoot{
Each columns show
(1) the observing epoch,
(2) the $T_{\rm B}$ ratio, and
(3) the magnetic field strength ratio.
}
\end{table}

\subsection{Estimation of the jet launching region size}\label{subsec:launching}

With high-resolution 86~GHz VLBI jet images, it is possible to estimate the diameter of the jet base $D_{0}$
by using the width and collimation profile of the outflow.
For this we assume a self-similar jet and that the observed power-law dependence can be used to 
back-extrapolate the jet width to its origin near the event horizon.
We then determine the collimation profile to $z=1R_{\rm sch}$, keeping the unknown separation $\epsilon$ 
between the BH and the VLBI core at 86\,GHz as a free parameter ($\epsilon\leq41\mu$as; see Sect. \ref{sec:ridge_analysis} and Figure \ref{fig:illust_geometry}).

\begin{figure}[t]
\centering
\includegraphics[width=8.5cm]{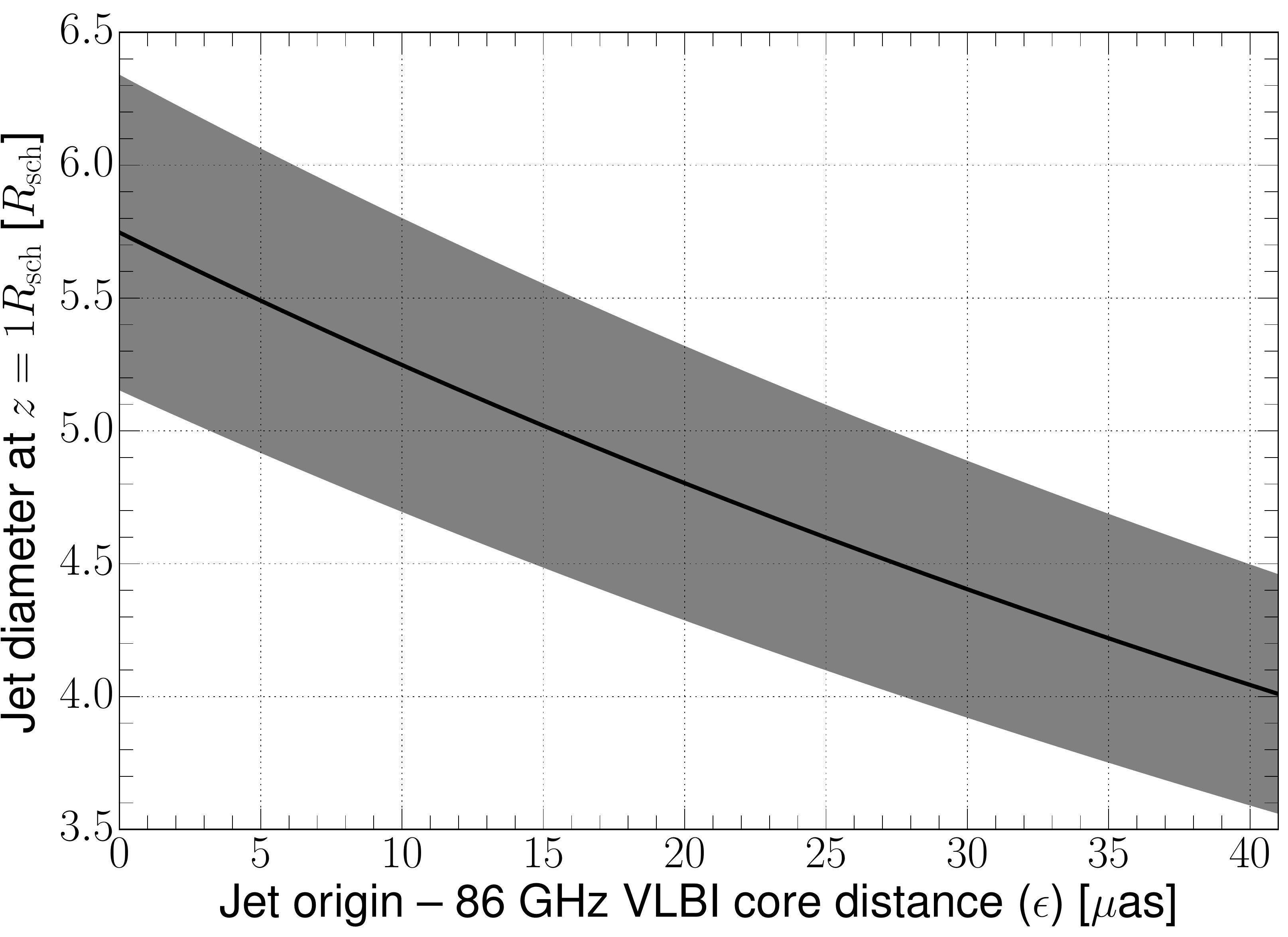}
\caption{
Diameter of the jet base at a distance of 1$R_{\rm sch}$ from the BH center
as a function of positional offset $\epsilon$ between BH and the 86\,GHz VLBI core (larger separation for larger $\epsilon$).
The solid line marks the diameter (width) of the jet base
and the shaded area denotes the $1\sigma$ uncertainty from the fitting.
}
\label{fig:dia_1rs}
\end{figure}

In Fig. \ref{fig:dia_1rs} we show the estimated size of the jet base 
versus the displacement $\epsilon$ of the VLBI core from the BH.
The width of the jet base is in the range of $\sim (4.0-5.5) R_{\rm sch}$, 
with a statistical uncertainty of $\sim0.5R_{\rm sch}$.
This small size is consistent with the upper limit given by the 86\,GHz VLBI core FWHM size, which is $\sim11R_{\rm sch}$.
It is also consistent with a circular Gaussian modelfit size estimate of $R_{0}=(5.6\pm0.4)R_{\rm sch}$
(or equivalently $40\pm$1.8\,$\mu$as) obtained at 230\,GHz with EHT observations \citep{doeleman12},
although the 230\,GHz size measurements can be model-dependent due to the limited $(u,v)$-coverage in the early EHT experiments.

Physically, this jet base size is comparable to the diameter of 
the innermost stable circular orbit (ISCO) for a non-spinning BH 
$D_{\rm ISCO} \sim 6R_{\rm sch}$
and it is much larger than $D_{\rm ISCO}$ for a maximally spinning BH with prograde disk rotation 
($D_{\rm ISCO}=1R_{\rm sch}$).
On the other hand, the diameter of the ergosphere in a rotating BH ranges between $(1-2)R_{\rm sch}$ \citep{komissarov12}.
This suggests that 
the diameter of the jet base of M\,87 matches
the dimensions of the innermost portion of the accretion disk.
It is interesting to note that for some other nearby AGN-jets 
(e.g., 3C\,84, \citealt{giovannini18}; Cygnus A, \citealt{boccardi16_86g}) 
the jet base appears to be much wider (100s of $R_{\rm sch}$).

We also highlight a somewhat larger value of $D_{0} \simeq (9.6\pm1.6)R_{\rm sch}$ which \cite{mertens16} 
determined for M\,87 at 43\,GHz via an independent analysis.
The difference in values for the jet base at 86 and 43\,GHz might be simply due to systematics, 
but several other physical explanations are also possible.
One possibility is that multi-frequency VLBI observations see radio emission coming from different layers of 
a transversely stratified jet, i.e., with stronger emission from the outer layers at lower frequencies.
For instance, we refer to VLBI observations of Cygnus~A at 5\,GHz and 86\,GHz, which show significantly different 
jet transverse widths \citep{carilli91,boccardi16_86g}.

An alternate interpretation of these differing values at 86 and 43\,GHz could also be attributed to the jet viewing angle.
The jet of M\,87 has a small viewing angle of $15^{\circ}-30^{\circ}$. 
A transverse velocity stratification of the layered jet sheath would lead to differential Doppler boosting (see \citealt{komissarov90}).
This effect depends on the viewing angle and could lead to different jet widths at different frequencies.

The accretion disk geometry is another factor.
The accretion flow in the center of M\,87 is believed to be geometrically thick 
and the scale height of the jet launching point could also be model-dependent (see \citealt{yuan14} for a review).

It is important to note that the geometrically wide origin of the sheath 
does not exclude the existence of 
a more narrow relativistic spine which may be rooted in the BH ergosphere \citep{bz77}.
An apparently faint, Doppler-deboosted ultra-relativistic spine located at the center of the jet
could explain the observed limb-brightening (e.g., \citealt{komissarov90}).
The high-energy emission observed from M\,87 also favors the presence of a fast spine surrounded by the slower sheath 
(see discussions in \citealt{abramowski12}).
For instance, more recent measurements by the EHT with longer baselines \citep{krichbaum14} and 
theoretical modeling of the jet base region \citep{kino15} provide hints of 
the existence of a compact emission region significantly smaller than 
originally inferred from EHT observations with less extended baseline coverage \citep{doeleman12,akiyama15}.
We also point to complex transverse jet intensity profiles shown by recent deep imaging of M\,87,
which may be consistent with the idea of the multiple jet layers with different origins \citep{asada16,hada17}.

\subsection{Implication of the large intrinsic opening angle in the innermost region}\label{subsec:initial_opening}

In the jet acceleration and collimation zone,
the relationship between the jet opening angle and the jet speed is important.
In particular, \cite{hada16} suggested that %based on the observed shape of the jet that 
several reconfinement nodes may form or a jet breakout from a dense atmosphere might occur near the base of the jet in M\,87.
If this is the case, the expansion and acceleration of the jet will be significantly different from the collimation acceleration scenario
\citep{komissarov07,lyubarsky09}.
For instance, for a sufficiently narrow jet of a given speed the jet-crossing time can be 
short enough for pressure disturbances from the ambient medium to propagate across the jet (i.e., {\it causally connected}).
On the other hand, too wide an opening angle makes the jet less sensitive to the ambient pressure. 
The latter case could lead to a different jet acceleration mechanism
such as the rarefaction acceleration (e.g., \citealt{tchekhovskoy10,komissarov10}).

We examine the relationship between the intrinsic opening angle $\phi_{\rm int}$ and the bulk Lorentz factor $\Gamma$
at the core separation $d=60~\mu$as ($27R_{\rm sch}$).
According to \cite{komissarov09}, a causally connected jet should satisfy
\begin{equation}
\Gamma\phi_{\rm int}/2\lesssim \sigma^{1/2}
\label{eq:causal}
\end{equation}
where the factor 2 accounts for the half opening-angle and 
the $\sigma$ is the level of jet magnetization defined by the ratio of the Poynting flux to kinetic energy (i.e., $\sigma=1$ for equipartition).
For the jet viewing angle of $\theta=18^{\circ}-30^{\circ}$, the jet will be causally connected if
$\Gamma\lesssim2\sigma^{1/2}/\phi_{\rm int}=(1.3-1.8)\sigma^{1/2}$.
If we presume $\sigma\sim4$ at the core distance $d\sim60~\mu$as (jet radius $\sim0.15$\,mas; see Fig. 18 of \citealt{mertens16}),
we obtain upper limits on the apparent jet speed of $\sim(2.3-3.4)c$.
Previous observations, in contrast, suggest apparent speeds of $\sim0.5c$ in this region \citep{hada16,mertens16}
and corresponding $\Gamma\phi_{\rm int}/2\sim0.7-0.9$, which satisfies Eq. \ref{eq:causal}.

Therefore, our $\Gamma\phi_{\rm int}/2\lesssim1$ hints at a gradual collimation and acceleration of the jet base
instead of a sudden acceleration accompanied by a jet breakout from a dense atmosphere, 
similar to GRB jet acceleration which has an order of magnitude higher $\Gamma\phi_{\rm int}/2$ \citep{panaitescu02}.
It is also interesting to note that $\Gamma\phi_{\rm int}/2\sim0.7-0.9$ in the M\,87 jet base is significantly larger than
$\Gamma\phi_{\rm int}/2\sim0.1-0.2$ found in typical pc-scale jet systems \citep{jorstad05,pushkarev09,clausen-brown13}.
This implies $\Gamma\phi_{\rm int}$ is not constant for the whole AGN jet population.

It is also crucial to ascertain weather the wide expansion in the jet base can suppress different types of instabilities or not,
in particular the current-driven kink instability (CDI).
The CDI plays a critical role in determining the Poynting flux to kinetic energy conversion in high magnetization jet environments 
(e.g., \citealt{singh16}).
In order for an instability to propagate across the jet,
the jet expansion timescale should be longer than the timescale needed for an instability to 
propagate across the jet cross-section.
This leads to the following criterion:
\begin{equation}
 \frac{t_{\rm dyn}}{t_{\rm exp}} \approx
 \frac{\beta\Gamma\phi_{\rm int}/2}{\beta_{s}} \lesssim 1
 \label{eq:inst_crit}
\end{equation}
\citep{komissarov09} 
where 
$t_{\rm dyn}=(\Gamma\phi_{\rm int}z/2)/(\beta_{s}c)$ is the jet crossing timescale for the instability,
$\beta_{s}$ is the signal speed in units of $c$,
$t_{\rm exp}=z/(\beta c)$ is the jet expansion timescale,
and $\beta$ is the speed of the jet in units of $c$.
If the magnetic field is dynamically important, 
$\beta_{s}$ is azimuthal component of the Alfv\'{e}n speed, which is $\sim1$ for a highly magnetized plasma \citep{giannios06}.
By using numbers corresponding to the highly magnetized M\,87 jet base, 
we find that the condition in Eq. \ref{eq:inst_crit} is satisfied for the jet base.
This suggests that current-driven kink instabilities can survive the jet expansion and 
could be important in terms of the dynamics and energy conversion occurring within the jet.

\begin{figure}[t]
\centering
\includegraphics[width=8.5cm]{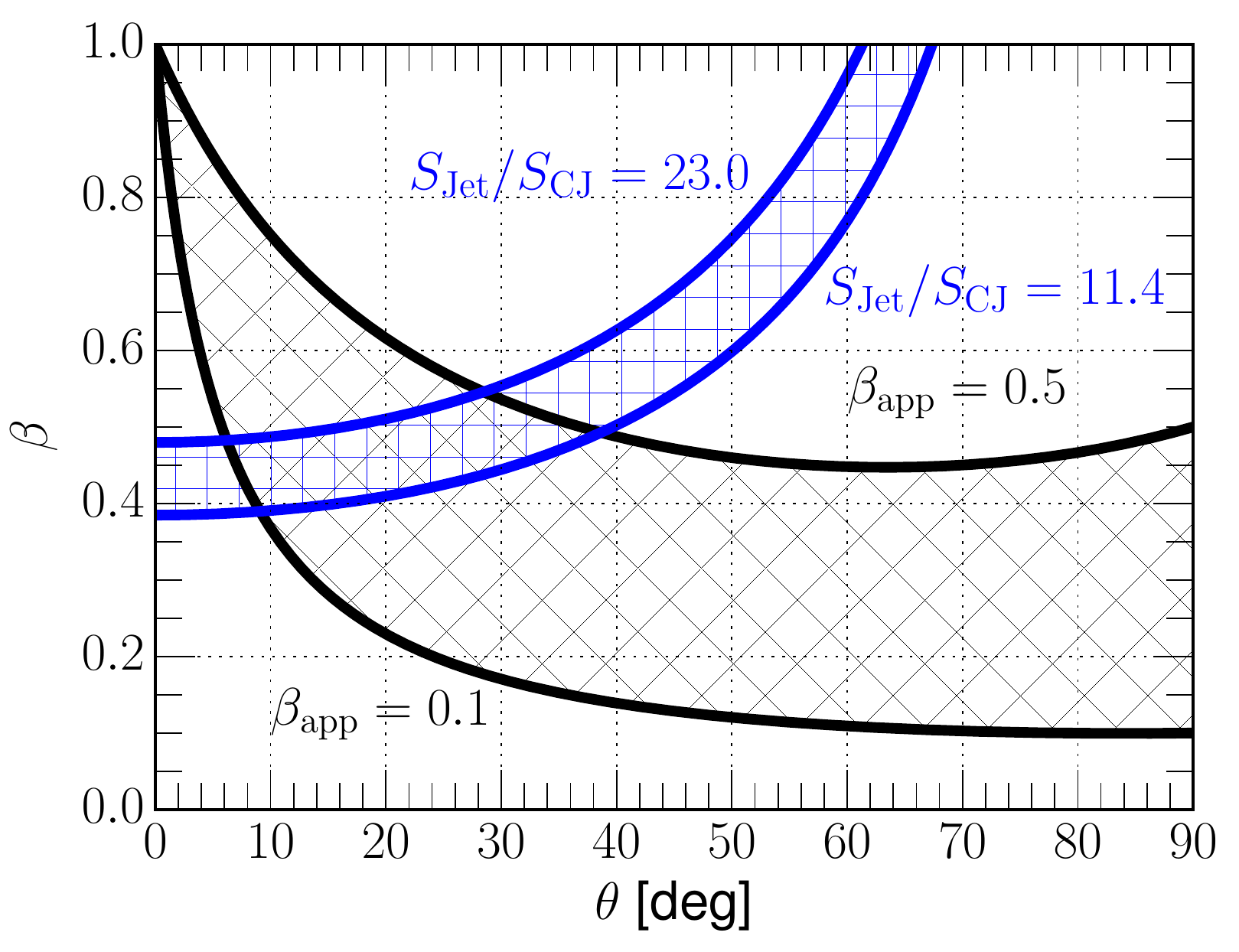}
\caption{
Possible range of M\,87 jet viewing angles and jet speeds (in units of $c$) 
that satisfy the measured jet to counter-jet ratio $R$ and our assumptions of the value of $\beta_{\rm app}$ 
at the core separation of $0.2-0.5$\,mas.
The cross and vertical hatches show allowed ranges of the parameters for $R$ and $\beta_{\rm app}$.
Only the overlapping region is physically allowed under the assumptions outlined in Sect. \ref{sec:CJ_Angle}.
}
\label{fig:ThetaBeta}
\end{figure}

\subsection{The inner jet viewing angle and the outflow speed}\label{sec:CJ_Angle}

The viewing angle and, more importantly, the velocity of the innermost M\,87 jet are crucial for our understanding of 
the jet geometry and dynamics.
Recent GRMHD simulations show that the filamentary jet structure becomes significantly complicated 
near the jet formation region (e.g., \citealt{moscibrodzka16}).
This can potentially make it difficult to determine precisely the viewing angle and the jet velocity.
Nevertheless, previous studies (e.g., \citealt{mertens16,walker16}) show that 
the two-dimensional kinematics of the jet in M\,87 can be decomposed into longitudinal and transverse motions.
The longitudinal motions can explain the majority of the relativistic boosting without additional jet curvature.
In the light of this result, we can test two possible scenarios using the observed jet to counter-jet ratio.
That is, the inner jet of M\,87 could have
(1) a stationary non-accelerating flow, with a constant jet speed and line of sight jet orientation (i.e., constant viewing angle)
or
(2) the viewing angle is constant but the jet is accelerating.

As for the case (1), if the approaching and receding jets are intrinsically of the same brightness and speed
and there is no transverse velocity gradient in both jets, 
the jet to counter-jet brightness ratio $R$ can be expressed by
\begin{eqnarray}
R & \equiv & I_{\rm Jet}/I_{\rm CJ}= \left(\frac{1+\beta\cos\theta}{1-\beta\cos\theta}\right)^{2-\alpha}
\label{eq:ratio}
\\
\beta & = & \frac{ \beta_{\rm app} }{ \sin\theta+\beta_{\rm app}\cos\theta }
\label{eq:bapp_in_bint}
\end{eqnarray}
where
  $I_{\rm Jet}$ and $I_{\rm CJ}$ are the intensities of the jet and the counter-jet,
  $\beta$ is the intrinsic jet speed normalized by the speed of light $c$,
  $\alpha$ is the spectral index ($S\propto\nu^{+\alpha}$), and
  $\beta_{\rm app}$ is the observed apparent speed.

It is worth noting that Eq. \ref{eq:ratio} assumes no transverse velocity gradient across the jet.
A faster jet speed closer to the central axis could make the apparent jet to counter-jet ratio $I_{\rm Jet}/I_{\rm CJ}$
lower because of different levels of Doppler boosting in the approaching and receding flows
(see Fig.2 and 3 of \citealt{komissarov90}).
We note that kinematics studies of the inner jet of M\,87 find
the same intrinsic flow speeds in the boundary layers of the jet and counter-jet within observational uncertainties \citep{mertens16}.
Hence, we do not include the transverse velocity gradient effect in the analysis of the jet to counter-jet ratio.

In order to find a range of $\theta$ and $\beta$ satisfying both Eq. \ref{eq:ratio} and \ref{eq:bapp_in_bint},
we assume $\beta_{\rm app}\sim0.5$ and $\sim0.1$ as the upper and the lower limits on the apparent speed, respectively \citep{hada16,mertens16}.
This wide range represents the variety of the directly measured jet speeds.
We also adopt a typical optically thin spectral index of the M\,87 jet $\alpha=-1$ \citep{hovatta14}. 
The jet to counter-jet ratio value is taken from Sect. \ref{subsec:cj} ($17\pm6$).
In Fig. \ref{fig:ThetaBeta} we show the possible range of $\theta$ and $\beta$.
The range of the possible jet viewing angle is wide ($6^{\circ}-38^{\circ}$), 
mainly because of the large scatter in the apparent jet speed.
If the fast speed ($\sim0.5c$) is directly associated with the true jet flow \citep{mertens16},
rather large jet viewing angles of $28^{\circ}-38^{\circ}$ are expected (c.f., \citealt{hada16}).
On the other hand, recent observations found significant acceleration within the jet of M\,87 \citep{mertens16,walker16}, 
which disfavor case (1).
Therefore, we expand our analysis and put more emphasis on case (2).
In this subsequent analysis we assume a viewing angle of $18^{\circ}$ \citep{mertens16},
which better explains fast superluminal motions in the outer jet.
If the jet to counter-jet ratio $R$ is given as a function of the distance from the jet origin,
the jet kinematic parameters including
the intrinsic speed $\beta$, 
the apparent speed $\beta_{\rm app}$, 
the bulk Lorentz factor $\Gamma$, 
and the Doppler factor $\delta$ can be computed along the jet axis via:
\begin{eqnarray}
 \beta &=& \frac{1}{\cos\theta}\left( \frac{R^{1/(2-\alpha)}-1}{R^{1/(2-\alpha)}+1}\right)~,
 \label{eq:bint}
 \\
 \beta_{\rm app} &=& \frac{\beta\sin\theta}{1-\beta\cos\theta} %~,
 \label{eq:bapp}
 \\
 \Gamma &=& \frac{1}{\sqrt{1-\beta^{2}}} %~,
 \label{eq:gamma}
 %\\
\end{eqnarray}

\begin{figure}[t]
\centering
\includegraphics[width=0.40\textwidth]{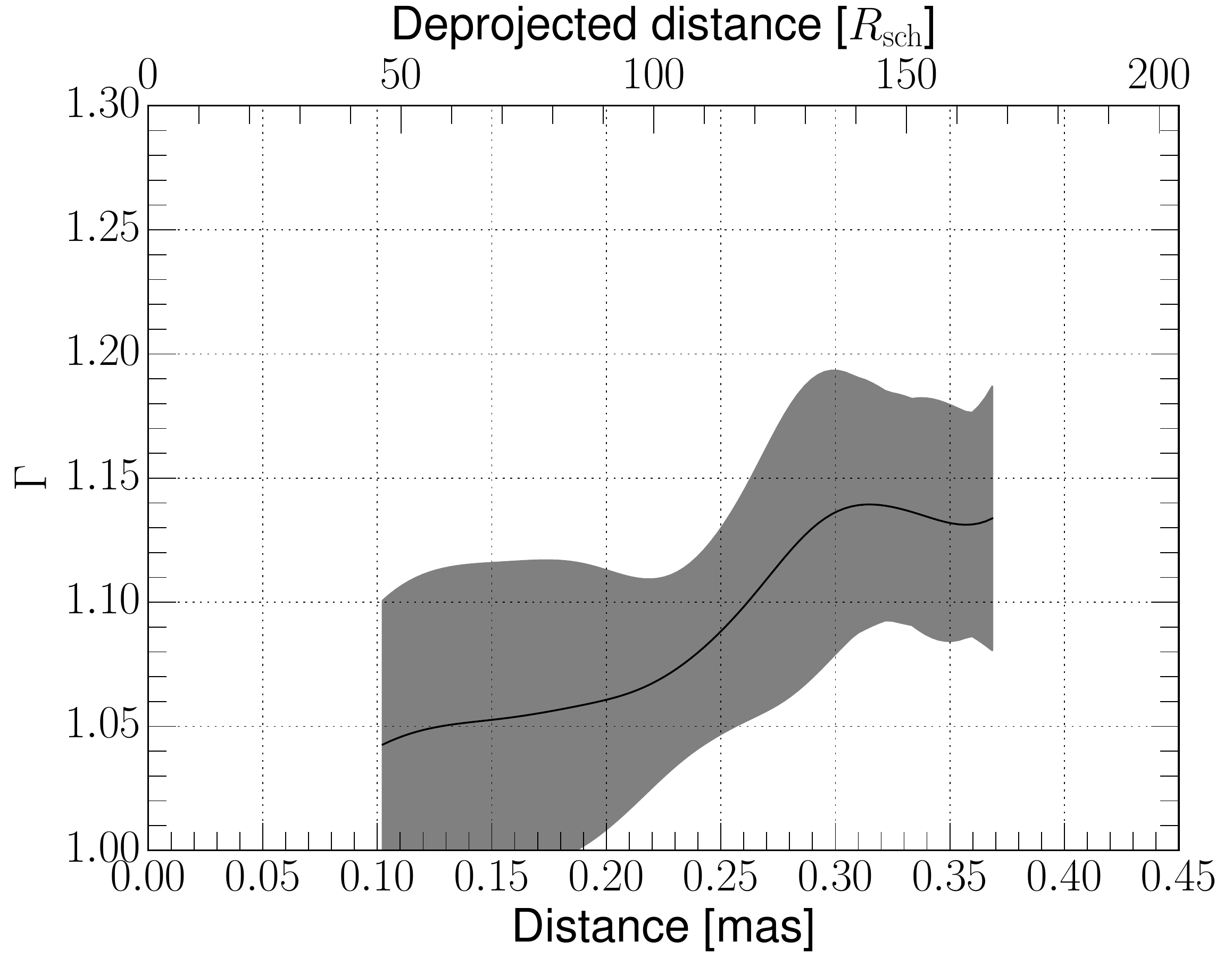}
\caption{
The bulk Lorentz factor of the inner jet of M\,87
estimated from the jet to counter-jet ratio versus the distance from the central engine 
under the assumption of stationary viewing and changing flow speed. 
The solid line and the shaded region indicate the mean values and the uncertainties, respectively.
The deprojected distnace has been calculated based on the viewing angle of $18^{\circ}$.
}
\label{fig:acceleration}
\end{figure}

In Fig. \ref{fig:acceleration} we show the results of our calculations of the Lorenz factor $\Gamma$.
In this model, the Lorentz factor mildly increases from $\sim1.0-1.1$ at 0.1~mas to $\sim1.1-1.2$ at $\sim0.35$~mas, which suggests mild inner jet acceleration.
In terms of the apparent speed, our fiducial model predicts $\beta_{\rm app} \approx (0.1-0.7)$ at 0.1\,mas and 
$(0.6-1.0)$ at $>0.3$\,mas.
The latter is comparable to what has been directly measured from VLBA 43~GHz observations at similar and/or slightly
larger distances (see Fig. 16 of \citealt{mertens16}).
We also recall that no clear inter-day timescale structural variation was seen in the jet in the 2009 data.
This implies an upper limit on the apparent jet speed of $0.2\times50~\mu{\rm as}/1~{\rm day}\sim3.65~{\rm mas}/{\rm yr}\sim1~c$
(using the speed conversion factor $1~c = 3.89~{\rm mas/yr}$ for M\,87 and adopting 1/5th of the beam size as the resolution limit).
The inferred range of the apparent jet velocity agrees well with this upper limit.

Hence, we conclude that a mildly accelerating inner jet model with a stationary viewing angle 
could explain the observed inner jet to counter-jet ratio.

\subsection{The spine-sheath scenario in the M\,87 jet}\label{sec:spine}

The transverse intensity profiles presented in Sect. \ref{subsec:slice}, Fig. \ref{fig:stack}c, and Fig. \ref{fig:slice_avg} 
suggest the existence of a central emission lane close to the jet base.
Other deep imaging experiments also suggest similar complex structure at more distant regions along the jet 
(\citealt{mertens16,asada16,hada17}; see also Sect. \ref{subsec:launching}).

\begin{figure}[t]
\centering
\includegraphics[width=8.5cm]{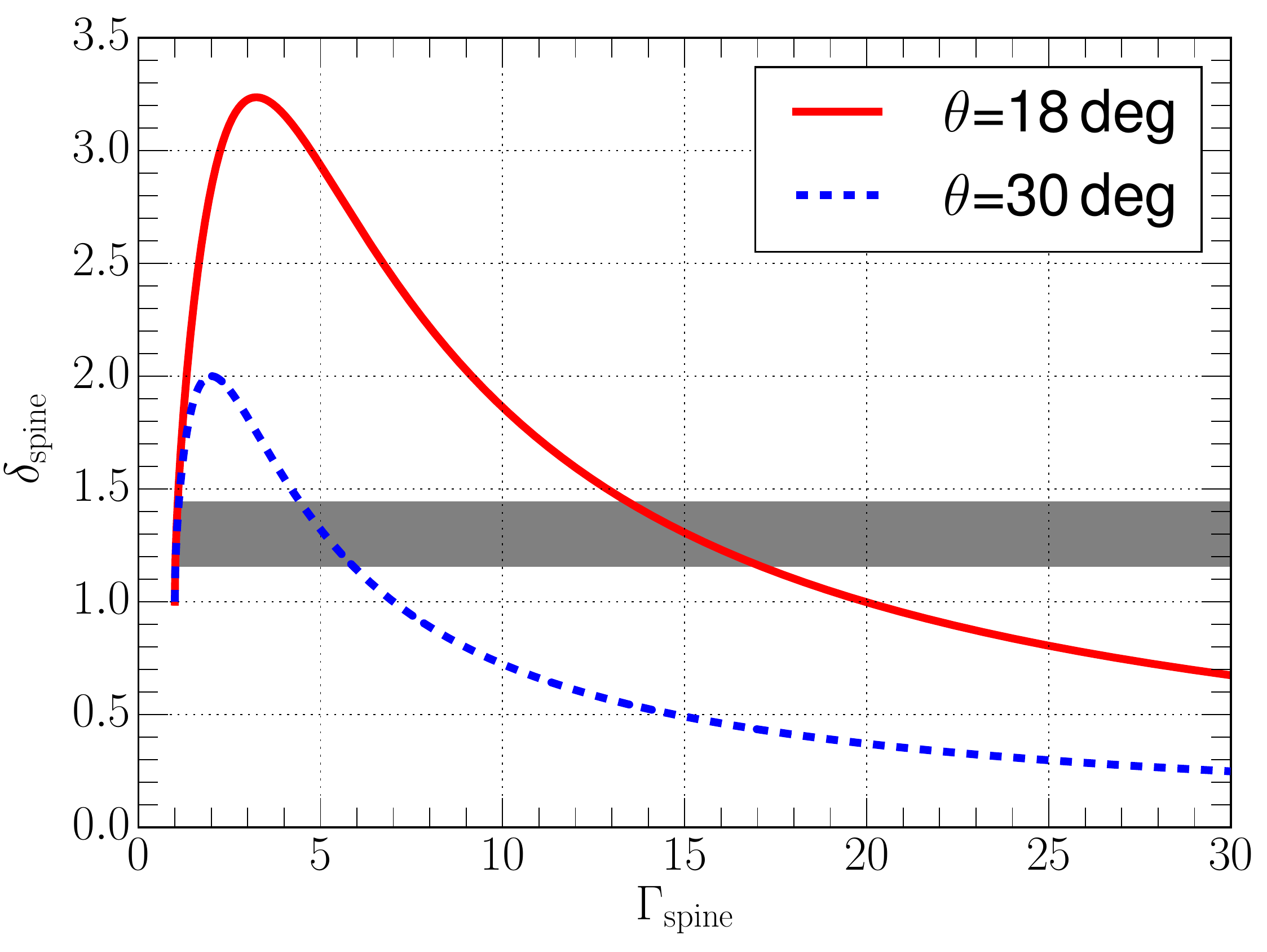}
\caption{
Doppler factor versus Lorentz factor for the spine.
The shaded region indicates possible range of the $\delta_{\rm spine}$ 
derived based on the assumption of the velocity stratification.
}
\label{fig:gamma_delta}
\end{figure}

If the central lane appears fainter only due to the Doppler de-boosting,
then the center-to-limb brightness ratio $\rho_{CL}$ 
measured at the core distances 0.5-1.0\,mas
can be related to the ratio of their Doppler factors by 
$\rho_{\rm CL} = (\delta_{\rm spine}/\delta_{\rm sheath})^{2-\alpha}$~. 
To calculate the expected range of the Spine Lorentz factor $\Gamma_{\rm spine}$,
we assume $\alpha=-1$, a viewing angle of $18^{\circ}-30^{\circ}$, and an apparent speed of the sheath of $\sim0.5c$
(the corresponding $\Gamma_{\rm sheath}\sim1.18-1.29$ and $\delta_{\rm sheath}\sim1.58-1.96$).
The observed $\rho_{\rm CL}$ constrains $\delta_{\rm spine}$ to be $\sim1.16-1.44$.
In Fig. \ref{fig:gamma_delta} we show the range of $\delta_{\rm spine}$ versus 
the spine Lorentz factor for small and relatively large viewing angles, respectively.
It is apparent from the figure that
a rather fast $\Gamma_{\rm spine}\sim13-17$ ($\Gamma_{\rm spine}\sim4.4-5.9$) 
is required
\footnote{
Here we exclude the apparent solution, i.e. $\Gamma_{\rm spine}\sim1$.
}
when the viewing angle is $18^{\circ}$ ($30^{\circ}$).

We note that the inferred $\Gamma_{\rm spine}$ is large considering the very small distance from the core 
($\sim0.5-1.0$\,mas; $70-140R_{\rm sch}$ projected).
For instance, $\Gamma_{\rm spine}\sim13-17$ is comparable to the Lorentz factor of HST-1 
determined at optical wavelengths by \cite{biretta99} ($\Gamma\sim14$ for the viewing angle $18^{\circ}$).
At such short wavelengths the observed radiation presumably traces relativistic plasma 
closer to the central axis of the jet (e.g., \citealt{perlman99,mertens16}).

GRMHD simulations show that accreting BH systems can form
(i) a narrow, relativistic, and Poynting-dominated beam (i.e., spine) and 
(ii) broader, sub-relativistic, and mass-dominated outflow (i.e., sheath) 
as a natural consequence of the mass accretion and BH physics (e.g., \citealt{hawley06,sadowski13}).
The latter helps to main low density levels near the central axis of the jet.
In such cases, the narrow beam propagates without significant mass loading
and maintains its initial speed up to large distances.

On the other hand, \cite{asada16} showed that the central lane expands in a similar manner as the sheath.
If the collimation and acceleration pattern of the spine is qualitatively similar to that of the sheath 
\citep{asada12,hada13,asada14,mertens16}, the Lorentz factor of the spine may increase at larger distances.
If this is the case, the smaller $\Gamma_{\rm spine}\sim4-6$, 
inferred assuming a larger inner jet viewing angle ($30^{\circ}$), is perhaps a more plausible estimate of the intrinsic speed.

It should be noted, however, that the intrinsic synchrotron emissivity of the spine and the sheath 
are not necessarily the same, especially when the two different layers are launched from different origins
(e.g., the spine from the BH while the sheath from the inner disk).
Closer to the central engine, the intrinsic emissivity of the spine may decrease \citep{mertens16}.
In addition, the central lane is comparably or even brighter than the limb at 1.6-5\,GHz (\citealt{asada16}; see Fig. 3 and 5 therein).
Such an intensity profile cannot solely be produced by a transverse velocity gradient.
Therefore, the limb-brightened jet morphology close to the jet base may not be fully explained by the velocity stratification.
Intrinsic differences in the jet plasma such as the jet composition and/or the magnetic field strength 
would need to be considered for the different jet emissivity.

\subsection{Evolution of the jet within $\lesssim100R_{\rm sch}$ de-projected distances}\label{sec:discuss_shape}

The commonly invoked ``differential collimation'' process (see Sect. 4.3.2 of \citealt{komissarov12}) 
requires significant magnetic hoop stress from toroidal magnetic fields.
However, this condition can be satisfied only in the ``far-zone'' where 
the toroidal field is substantially stronger than the poloidal field
(e.g., \citealt{komissarov07,komissarov09,lyubarsky09}).
Therefore, a magnetic hoop stress might not be valid near the jet launching region where 
the poloidal field components are much more dominant (e.g., \citealt{tchekhovskoy15}).
For this scenario, we can estimate the radial distance of such a critical point from the central engine.
A substantial change in the B-field orientation occurs when the jet radius approaches the light cylinder radius
$r_{\rm lc}=c/\Omega$ \citep{meier12} where $\Omega$ is the angular speed of the outflow.
\cite{mertens16} recently determined the light cylinder $r_{\rm lc}$ in the M\,87 jet to be $\sim20R_{\rm sch}$.
The observed jet radius becomes comparable to $r_{\rm lc}$ at $\sim0.2$\,mas ($28R_{\rm sch}$ projected) core separation.
It is interesting to note that we detect a slight divergence between the data and the single power-law model 
at a similar core separation (see the bottom panel of Fig. \ref{fig:collimation}).
Future higher angular resolution imaging experiments with 
the EHT plus a
phased Atacama Large Millimeter/submillimeter Array (ALMA)
\citep{matthews17}
will be able to reveal the structure and propagation of the M\,87 jet on such spatial scales.

\section{Conclusion}\label{sec:conclusion}

In this paper, we present a study of 
the physical conditions and structure of the innermost jet in M\,87 at  projected distances of 
$(7-100)~R_{\rm sch}$ from the core.
We summarize our findings and conclusions as follows:

\begin{enumerate}

\item

Deep images of the jet base in M\,87 were obtained with an east-west spatial resolution of 
$7R_{\rm sch}$ using the GMVA at 86~GHz.
The multi-epoch images obtained over a decade show consistency in the core-jet structure including 
a highly limb-brightened jet, a faint central lane between the edges of the jet, and a weak counter-jet 
(but no significant evidence for counter-jet limb-brightening). 

\item

The VLBI core has a mean resolved size of $\sim11R_{\rm sch}$ at 86~GHz.
The apparent brightness temperature of this compact region is $\sim(1-3)\times10^{10}~K$,
nearly an order of magnitude lower than the equipartition brightness temperature ($\sim2\times10^{11}~K$).
This implies that the core is magnetically energy dominated.
The corresponding magnetic field strength would be between $61-210$~G
for an equipartition magnetic field strength of $\sim1$~G in the VLBI core region.

\item

The size of the jet launching zone is estimated assuming a self-similar jet structure.
We find the diameter of the jet base of $\sim4.0-5.5R_{\rm sch}$, which points to the inner accretion disk as the origin of the sheath.
This size is in good agreement with the upper limit set by 86~GHz VLBI core size measurements ($11R_{\rm sch}$) and
results from previous EHT observations \citep{doeleman12,akiyama15}.

\item

The apparent opening angle of the jet base can be as large as $127^{\circ}\pm22^{\circ}$ 
at 60$~\mu$as core separation (27$R_{\rm sch}$ de-projected).
Combining the jet geometry and the kinematic information, we find that 
the jet base is still causally connected
despite the unusually wide opening angle ($\Gamma \phi_{\rm int}/2 = 0.7-0.9$).
On the other hand, a time scale analysis shows that the jet base region can be sensitive to 
the current driven instabilities in the strong magnetic field regime.

\item 

We considered two distinct jet models; one in which the viewing angle and the speed of the jet was fixed,
and the other in which the viewing angle was fixed but the jet speed varies with distance from the central engine.
We find the jet speeds ranging from $0.1-0.5c$ for the former and $0.1-1.0$c for the latter.

\item

We investigate the origin of the faint central emission lane at 86~GHz by first considering a 
pure transverse velocity stratification scenario 
(i.e., the spine-sheath model).
We constrained the Lorentz factor of the spine to be $\Gamma_{\rm spine}\sim13-17$ ($4-6$) for $\theta=18^{\circ}$ ($30^{\circ}$)
and compared it with the Lorentz factor of HST-1 determined at optical wavelengths ($\Gamma\sim14$).
These estimates are consistent with a nearly constant velocity of the spine ($\theta=18^{\circ}$) or 
an accelerating spine with rather large inner jet viewing angle ($\theta=30^{\circ}$).
However, we notice that the relative brightness of the central lane with respect to the edges become 
significantly larger at longer centimeter wavelengths.
Therefore, we suggest that the edge-brightening in the jet of M\,87 may not be driven only by the velocity gradient 
but also by intrinsic differences (i.e. composition, magnetic field strength) in the plasma
within the lane and the sheath.

\end{enumerate}

\begin{acknowledgements} 
% General
We sincerely thank the referee, Serguei S. Komissarov, 
for the review and comments which improved the interpretation of our results.
We also thank Nicholas R. MacDonald for the careful reading and helpful comments.
% Kim
J.-Y. Kim is supported for this research by the International Max-Planck Research School (IMPRS) for Astronomy and Astrophysics at the University of Bonn and Cologne.
% Ros
E.R. was partially supported by the Spanish MINECO grant AYA2015-63939-C2-2-P and by the Generalitat Valenciana grant PROMETEOII/2014/057.
% GMVA
This research has made use of data obtained with the Global Millimeter VLBI Array (GMVA), 
which consists of telescopes operated by the MPIfR, IRAM, Onsala, Metsahovi, Yebes, the Korean VLBI Network, the Green Bank Observatory and the Long Baseline Observatory (LBO). 
The VLBA is an instrument of the LBO, which is a facility of the National Science Foundation operated by Associated Universities, Inc. 
The data were correlated at the correlator of the MPIfR in Bonn, Germany.
% DiFX
This work made use of the Swinburne University of Technology software correlator, developed as part of the Australian Major National Research Facilities Programme and operated under licence.
\end{acknowledgements}


\bibliographystyle{aa}
\bibliography{m87bibliography.bib}

\end{document}